\documentclass[graybox, envcountchap]{svmult}
\usepackage{mathptmx}        % selects Times Roman as basic font
\usepackage{amsmath}
\usepackage{amssymb}
\usepackage{color}
\usepackage{helvet}          % selects Helvetica as sans-serif font
\usepackage{courier}         % selects Courier as typewriter font
\usepackage{dirtree}
\usepackage{makeidx}        % allows index generation
\usepackage{graphicx}        % standard LaTeX graphics tool
\usepackage{subfig}
\usepackage{multicol}        % used for the two-column index
\usepackage[bottom]{footmisc}% places footnotes at page bottom
\usepackage{hyperref}        %for hyperlinks
\hypersetup{colorlinks=true,urlcolor=blue}

\usepackage[misc]{ifsym}

\makeindex

\graphicspath{{figures/}}

\usepackage{listings}
\usepackage{xspace}
\usepackage{xfrac}

\usepackage[
  sorting=nyt,
  style=ieee
]{biblatex}
\newcommand{\bhname}{M87$^*$\xspace}
\newcommand{\sgra}{{Sagittarius~A$^{\ast}$}\xspace}
\def\sgr{Sgr~A$^{\ast}$\xspace}
\newcommand{\kharma}{{KHARMA}\xspace}
\newcommand{\iharm}{\texttt{iharm3D}\xspace}
\newcommand{\Rpeak}{R_\mathrm{peak}}

\newcommand{\nvidia}{{Nvidia}\xspace}

\begin{document}

\title{\kharma: Flexible, Portable Performance for GRMHD}
\author{Cora Prather}
\institute{Cora Prather (\Letter) \at Los Alamos National Laboratory, PO Box 1663, Los Alamos, NM}

% Usual enumerate, itemize, table/tabular
% paragraph emphasis \begin{svgraybox}
% Math envs: eqnarray, theorem, definition, proof w/ \qed

%Colored figures are welcome in any standard format (jpg, tif, ppt, gif). If possible please provide the original figure in high resolution (300 dpi minimum). {\color{red}\bf Please do not forget to obtain permission in case the figure is from a published article}. For journals like ApJ, PRD, PRL, etc. you do not need the permission if you are an author of the article of the original figures, but for most journals you need the permission even if you are the author of those figures.
% If the width of the figure is less than 7.8 cm use the \texttt{sidecapion} command to flush the caption on the left side of the page. If the figure is positioned at the top of the page, align the sidecaption with the top of the figure -- to achieve this you simply need to use the optional argument \texttt{[t]} with the \texttt{sidecaption} command
% scale w/[scale=.95]

\maketitle

%Each chapter should be preceded by an abstract (about 250 words) that summarizes the content. The abstract will not appear at the beginning of the chapter in the hard copy of the book. The abstract will appear only \textit{online} at \url{www.SpringerLink.com} and be available with unrestricted access. This allows unregistered users to read the abstract as a teaser for the complete chapter.
\abstract{
\kharma (an acronym for ``Kokkos-based High-Accuracy Relativistic Magnetohydrodynamics with Adaptive mesh refinement'') is a new open-source code for conducting general-relativistic magnetohydrodynamic simulations in stationary spacetimes, primarily of accretion systems.  It implements among other options the High-Accuracy Relativistic Magnetohydrodynamics (HARM) scheme, but is written from scratch in C++ with the Kokkos programming model in order to run efficiently on both CPUs and GPUs.  In addition to being fast, \kharma is written to be readable, modular, and extensible, separating functionality into ``packages,'' representing, e.g., algorithmic components or physics extensions.  Components of the core ideal GRMHD scheme can be swapped at runtime, and additional packages are included to simulate electron temperature evolution, viscous hydrodynamics, and for designing chained multi-scale ``bridged'' simulations.  This chapter presents the computational environment and requirements for \kharma, features and design which meet these requirements, and finally, validation and performance data.
}
\vspace{1cm}

\kharma\footnote{\url{https://github.com/AFD-Illinois/kharma}} (an acronym for ``Kokkos-based High-Accuracy Relativistic Magnetohydrodynamics with Adaptive mesh refinement'') is a new open-source performance-portable code for performing general-relativistic magnetohydrodynamic (GRMHD) simulations, targeting stationary spacetimes such as black hole accretion systems.  It implements the High-Accuracy Relativistic Magnetohydrodynamics (HARM) scheme \cite{gammie2003}, written from scratch in C++ 17 to leverage the Kokkos programming model \cite{trott2021} in order to run efficiently on various computer and accelerator architectures using the same source code.  \kharma also uses the Parthenon adaptive mesh refinement (AMR) framework \cite{grete2022} to support static and adaptive mesh refinement and to provide a consistent overall code structure.

\kharma is in production use as one of the primary codes used for GRMHD simulations in the Event Horizon Telescope Collaboration (EHTC), furnishing many of the GRMHD simulations used in official papers such as \cite{EHTSgrAPaper5,EHTSgrAPaper8} and collaboration projects such as \cite{ricarte2023a,emami2023}. In addition to being fast, \kharma has been designed from the ground up to be modular and extensible through the use of packages, designed to each contain a single algorithmic component or physics extension.  This modularity extends to all features, such as the magnetic field transport and primitive variable inversion, in addition to the usual modular components such as reconstruction scheme and Riemann solver.  Various additional physics packages have been developed to treat new systems and processes beyond ideal GRMHD, including an approximate viscosity treatment useful for very low accretion-rate systems as described in \cite{chandra2015}, and a co-evolved electron temperature treatment for two-temperature plasmas as described in \cite{ressler2015}, as well as a mode for running chained ``multi-scale'' simulations as described in \cite{cho2023}.  An extension is also being developed for simulating radiative effects, which are extremely important at high accretion rates and increasingly computationally feasible for GPU-based codes on modern supercomputers.

This chapter provides a brief case study of \kharma, beginning with the hardware trends and community requirements which led to its development, then covering the code's scheme, software design, and features, and ending with basic validation and performance data. 

\section{Necessity of performance portability}
\label{sec:computing}

Perhaps the single choice that most dictates \kharma's code structure is to be ``performance-portable,'' targeting efficient operation on general-purpose graphics processing units (GPUs) as well as other accelerator architectures and traditional central processing unit (CPU) cores.  Adding performance portability to an existing code often requires substantially re-writing it to leverage new loop ordering and data structures, and this difficulty was a primary motivator for developing \kharma from scratch.

In order to understand the trends in computing which made performance portability so important as to abandon an existing fast C implementation of HARM \cite{prather2021}, it's useful to provide a bit of history and context around high-performance computing (HPC).  This context is unnecessary for understanding \kharma itself; readers already convinced of the universal need for performance portability can skip to \kharma's more unique code requirements in Section \ref{sec:requirements}.

\subsection{Performance and power}
\label{sec:computing_chips}

In 1965, Gordon Moore predicted that the number of transistors on a given integrated circuit would double each year.  In 1975, he revised this to predict a slower doubling rate of 2 years from 1980 onward \cite{moore1965,moore1975}.  Whether by prescience or self-fulfilling prophecy, this doubling rate has held for nearly all of the ensuing 4\sfrac{1}{2} decades.  Usually, this increase is interpreted as computer chips getting ``faster'' or ``more capable.''

However, over the years, the purpose of these transistors -- that is, how new transistors are ``spent'' to improve a chip's capabilities, has shifted.  For the first decades of chip development, from the 1970s to 2000s, extra transistors were used to increase the clock speed and instruction processing rate of otherwise similar chips -- that is, to make existing programs run faster.  However, power constraints cut short the increasing clock rates in the mid-2000s, necessitating that designers find other ways to improve chips -- thus, improvements to CPUs since that time have taken the form of adding more processing cores to a single chip, operating at the same frequency but processing more data (so long as programs could take advantage of the parallelism).  This much is well-known in the supercomputing community.

Parallel to this more obvious trend of multiple cores, we see a trend toward not just duplicating existing processors, but specializing them for particular kinds of work -- seen in GPUs, as well as vector-optimized or ``many-core'' CPUs.  Specialization involves taking out functionality like branch prediction, out-of-order execution, and large instruction sets, and using their area and transistor count to further boost the amount of data a chip can process at once, through additional parallel execution units.  At the far end of specialization, chips capable of providing only one function (encryption, say, or video encoding) can provide orders of magnitude faster processing than software implementations.

\begin{figure}
    \centering
    \includegraphics[width=0.9\textwidth]{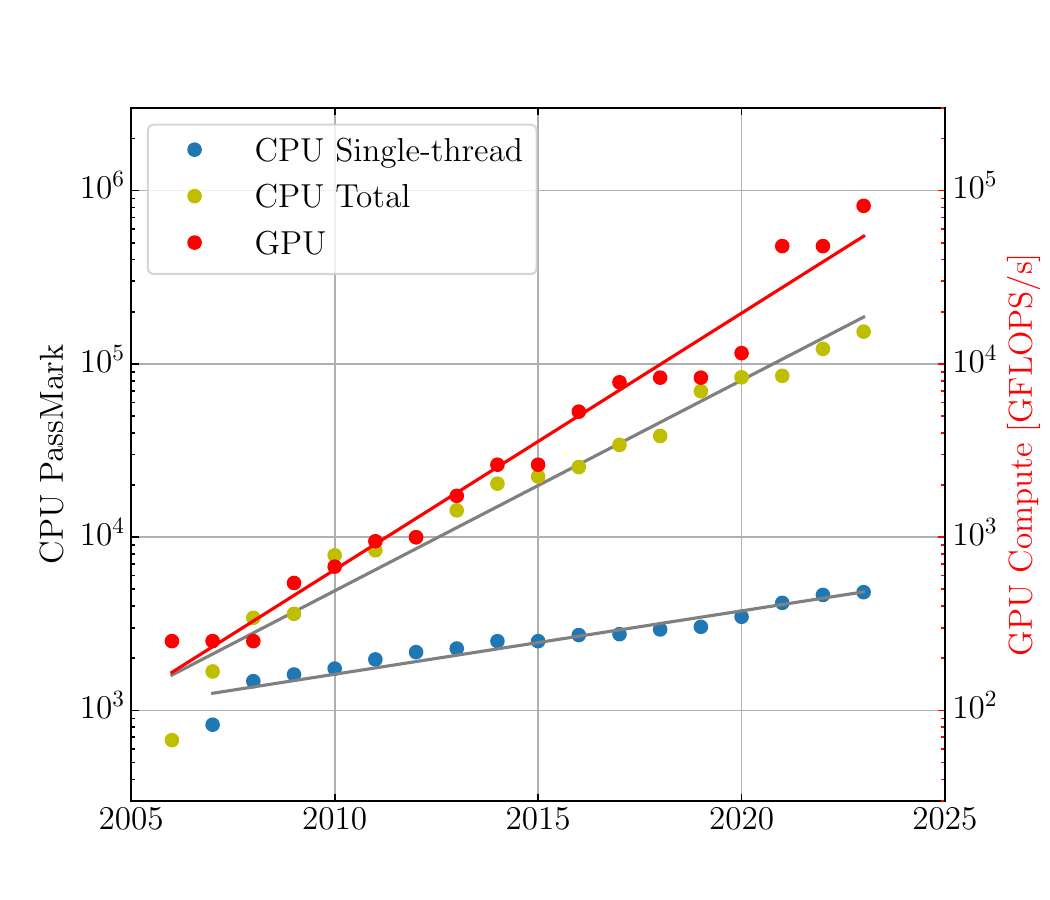}
    \caption{CPU single-core and total performance development, plotted against GPU theoretical peak performance performing floating-point operations.  Scales for CPU and GPU metrics are independent, and are adjusted so that metrics coincide in ~2006, but all scales span exactly 4 decades to accurately depict relative improvement. Fitting functions reflect exponential growth of $2^{0.12 t}$, $2^{0.40 t}$, and $2^{0.49 t}$ for CPU single, full, and GPU performance respectively, with $t$ measured in years.
    CPU performance is measured via public listed results of the PassMark combined CPU benchmark software.  GPU performance is taken from combinations of manufacturer-listed and reviewer-tested floating point operations per second (FLOPS/s) at double precision (not AI-specific packed operations marketed as ``matrix'' or ``tensor'' cores).  CPU scores from \cite{passmarksoftware2022}, GPU performance from \cite{techpowerup2024}, figure originally from \cite{prather2022}.}
    \label{fig:clocks_core}
\end{figure}

Both trends are illustrated in parallel in Figure \ref{fig:clocks_core}.  Plotted in blue and gold are CPU scores in the PassMark general benchmark\footnote{\url{https://www.passmark.com/}}, for single cores and full sockets \cite{passmarksoftware2022}.  Single-core performance stagnates with the power limits to clock rate in 2008, and has grown at a rate of only $2^{0.12 t}$ (where $t$ is in years) since then, mostly consisting of hard-won efficiency improvements in the rate of instructions cleared per clock cycle (IPC).
Instead, performance benefits are realized by increasing the core count; parallel workloads have still seen a $2^{0.40 t}$ improvement as core counts have increased over the last decade.
Plotted in red are GPU performance numbers in floating-point operations per second (FLOPS/s) \cite{techpowerup2024}.  Note this is a significantly different measurement from a PassMark score, evaluating GPUs' effectiveness at number crunching specifically, not general workloads (which they are generally unsuited for anyway).  This means that GPU performance is plotted on a \textit{different axis} from CPU performance.  However, both axes span exactly four decades, to compare relative improvement, where GPUs clearly have an advantage, especially in recent years.  Over the same measured span 2006-2023, GPUs grow in floating-point performance at $2^{0.49 t}$, paying forward the full benefits of their increasing transistor count.

It should not be surprising that special-purpose chips can improve at their intended task faster than general-purpose chips can improve at a broad range of tasks; they trade generality for efficiency.  As power constraints and the end of Moore's law force the use of more creative methods to improve performance, the pressure toward specialization will likely increase.

The benefits of specialization are nowhere more evident than in Artificial Intelligence (AI).  AI research now drives the development of its own class of specialized chips, designed even more narrowly than GPUs for processing the large matrix multiplications used in training and inference.  The clear efficiency benefits of specialized hardware have driven a rush of new architectures, such as Google's Tensor Processing Units \cite{jouppi2017}, Cerebras systems' Wafer-Scale Engine \cite{lauterbach2021} and Tesla's many-core ``Dojo'' architecture \cite{talpes2023}, which have become increasingly widespread. Even architectures which remain general in principle are gaining subsystems targeted at AI calculations specifically (\nvidia's Tensor Cores, Intel's AMX Engines, etc.).

\subsection{Specialization in supercomputing}
\label{sec:computing_top500}

\begin{figure}
    \centering
    \includegraphics[width=0.9\textwidth]{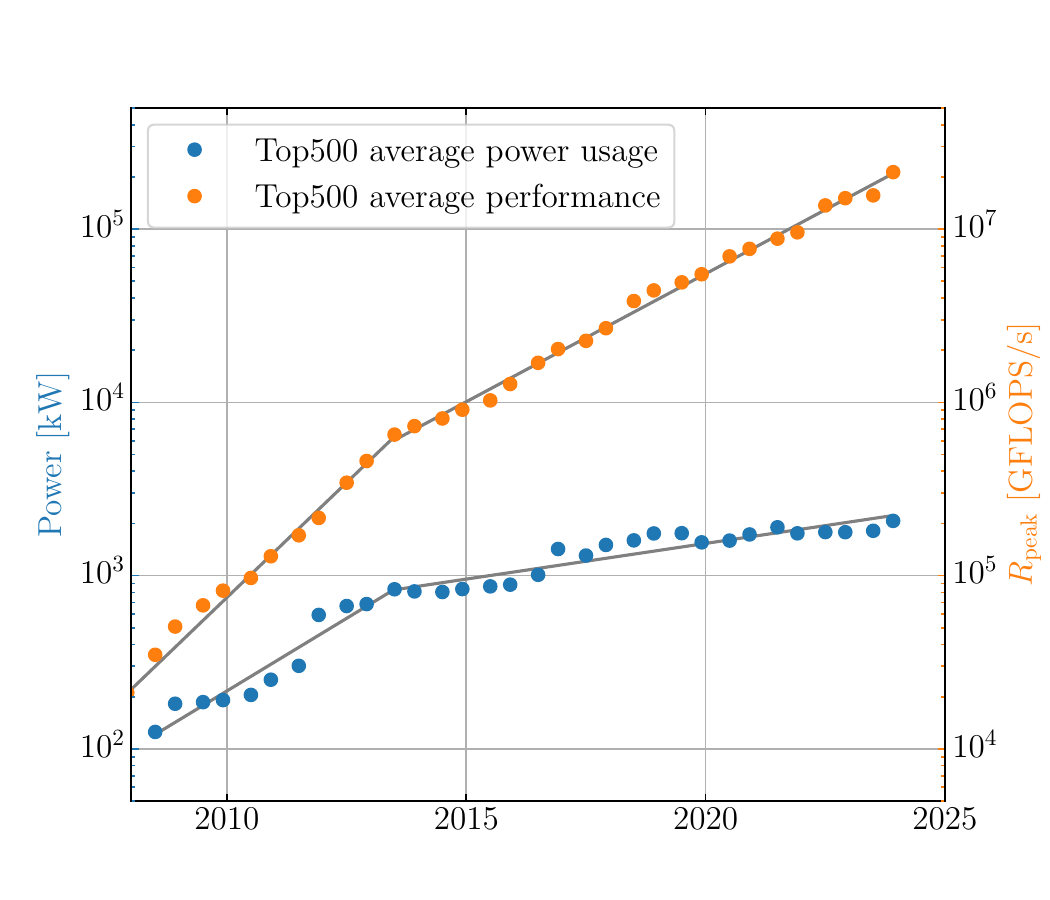}
    \caption{Average power draw of the top 500 machines, plotted with average performance development, showing a clear slowing of power usage increases in about 2013, when average machine power usage began to exceed 1MW.  Average power usage of the reporting fraction of the top 500 machines shows exponential growth at $2^{0.56 t}$ from 2008 (when reporting began) to 2013, and $2^{0.14 t}$ thereafter. Average peak performance $\Rpeak$ in GFLOPS/s of all top 500 machines shows development at $2^{0.88 t}$ before 2013, $2^{0.49 t}$ thereafter.
    Data from \cite{meuer2023}, figure originally from \cite{prather2022}.}
    \label{fig:top500_machine_power}
\end{figure}

Though more recently than the frequency limit, supercomputing has similarly hit a hard limit: power draw.  Machines consuming more than about 1MW can cost more in electricity than hardware, so the top machines cannot continue to draw more power without incurring spiraling costs.  To illustrate trends in supercomputing, we use the open top 500 supercomputers list compiled semi-annually \cite{meuer2023}.  While not a complete list of every machine, is still indicative of overall trends in supercomputers both public and private.  As shown at left of Figure \ref{fig:top500_machine_power}, from 2008 (when recording of power usage began) to 2013, the average power draw of the reporting portion of the top 500 machines grew by nearly an order of magnitude in 5 years.  Since then, however, it has grown by only another factor of 2, and appears to be slowing further still.  This is good news for global energy usage, of course, but it has visibly slowed the performance development of the top 500 machines, as shown at right of Figure \ref{fig:top500_machine_power}.  Before the increasing power usage slowed in 2013, machines doubled in performance every year; afterward, every two years.

Even discounting the very top machines, where exotic architectures have always been common, the closing power envelope is pushing HPC systems toward special-purpose chips, mostly GPUs.  Indeed, the number of machines in the top 500 list powered primarily by GPUs has increased substantially: in 2013, under 10\% of systems used some kind of GPU accelerator -- as of November 2023, the share is just about 30\%, together providing half of the list's total computing power \cite{meuer2023}\footnote{This number excludes specialized CPUs such as the Fujitsu A64FX and Intel Knights Landing architectures.  Nevertheless, these architectures present many of the same challenges and opportunities as accelerators.}.  While initial adoption of accelerators was slow due to programming challenges, maturing software (AI frameworks, programming models, etc..) and the closing power envelope have made GPUs an ever clearer choice for new systems.

Even the word GPU in the HPC context has shifted to mean something more specialized than it did a decade ago: back then, top machines such Oak Ridge National Laboratory's Titan were powered by what were effectively consumer graphics chips --  the GK110 chips used in that machine were also sold to consumers as the Titan GPU.  Even when the die was not exactly the same, the technologies were: many supercomputers over the course of the 2000s were built with technologies sold concurrently (or even primarily) to consumers.  Modern ``GPUs,'' in contrast, are specialized computational designs, increasingly distinct in architecture from their consumer counterparts (e.g.,~CDNA vs.~RDNA from AMD, XeHPC vs.~XeLP from Intel, Hopper and Blackwell vs.~Lovelace from \nvidia).

The sudden specialization has a clear cause: the massive hardware market for running artificial intelligence models.  As the large tech companies build AI training supercomputers on the scale of cloud server farms, they finally represent the kind of large and stable market that will drive stiff competition, and it seems likely that chip architectures will continue to shift to serve the needs of the larger AI market, rather than  traditional HPC use cases.  The current generation of supercomputers uses three substantially new architectures (Fujitsu A64FX, AMD CDNA2, Intel XeHPC), and as AI drives further architecture changes, the next generation may use substantially different architectures still.  Yanked out of its complacency by a strong AI hardware market and increasing power efficiency concerns, not to mention the end of Moore's law, HPC faces a decade of either rapid architectural changes or complete performance stagnation.

\subsection{Specialized hardware requires portable software}
\label{sec:computing_software}

It should be no surprise that in places where complex specialized hardware has found success, it has generally included an easy and standard software interface, to abstract away the complexity.  Take graphics acceleration: calls using a standard interface such as OpenGL or DirectX can be accelerated by a huge number of different hardware implementations, each providing a driver which translates the standard calls to specific instructions for the accelerator hardware.  Thus innovations to the hardware do not require changing existing applications at all.  A similar separation of standard interface layer and driver implementation layer has found success in enabling many other specialized hardware implementations, for video encoding/decoding, networking operations such as encryption, etc.

There is no single standard language of this kind for HPC as there is for other accelerators.  This is not for lack of trying, however, and there are promising modern contenders: projects within HPC such as Kokkos \cite{trott2021}, OCCA \cite{medina2014}, RAJA \cite{beckingsale2019}, as well as industry standards such as OpenCL\footnote{\url{https://www.khronos.org/opencl/}}, SYCL\footnote{\url{https://www.khronos.org/sycl/}}, OpenACC\footnote{\url{https://www.openacc.org/}}, and efforts within the C++ standard under the namespace \texttt{std::parallel}.  All of these attempt to express accelerator programming in a hardware-independent way, striking a balance between providing an easy-to-use interface and allowing low-level access and flexible programming.

While the lack of a standard is disappointing, it is not reason to avoid any of the existing options.  Nearly all of the listed performance portability options involve abstractions which simplify device programming, and provide convenient frameworks to simplify host programming as well.  Many of them share a similar overall structure for expressing compute kernels, so porting between them may often be easier than adapting ``landlocked'' pure C code of any complexity.

\section{Other Requirements}
\label{sec:requirements}

Besides performance portability, there were several requirements that shaped \kharma's design.  \kharma was designed primarily as a code for black hole accretion simulations, specifically of low-luminosity active galactic nuclei (LLAGN), that is, very low-accretion-rate supermassive black hole (SMBH) accretion systems where radiation does not play a primary role in the fluid dynamics.  The primary targets were \bhname and \sgra, in order to understand and utilize the images from the Event Horizon Telescope (EHT).  This provided a simple and useful target for early development, though more recently the code has been extended well beyond this original purpose.

The following sections give background on the scientific requirements which shaped \kharma's development.

\subsection{Simulation Libraries}
\label{sec:libraries}
The first black hole (BH) images of \bhname by the EHT Collaboration (EHTC) \cite{EHTM87Paper1} presented a unique opportunity to put accretion theories and simulations directly to the test, by comparing concrete predictions from ray-traced simulations to the EHT image properties and other measurements of \bhname.  Within the collaboration, a parameter study or ``library'' of GRMHD simulations was created.  The core of this library consisted of 10 simulations performed with the HARM-based GRMHD code \iharm \cite{prather2021}.  Simulations were performed on Stampede2 at the Texas Advanced Computing Center (TACC), where a set of just 10 simulations used a significant allocation of CPU-based resources.

From this core library of GRMHD simulations, images could be created via general-relativistic ray-traced volume rendering of the diffuse optically thin emission, generating a prediction for emission intensity at EHT frequencies \cite{wong2022}. With these images, especially when used in combination with other measurements (e.g., estimates of the jet power), the collaboration was able to significantly narrow the space of realistic accretion possibilities, and thereby understand the salient image features in terms of the accretion state around the event horizon (EH) \cite{EHTM87Paper5}.  Many research projects within the collaboration have made use of this standard library, from validation and calibration for official results \cite{EHTM87Paper4,EHTM87Paper6} to interpretation of polarized results \cite{EHTM87Paper8}, and many individual studies \cite{johnson2020,gold2020,palumbo2020,lin2020,ricarte2021}.

However, running libraries of simulations is expensive, and the conclusions are limited to motivating or demotivating particular models within a set of options necessarily limited in both its coverage of the available parameters, and its included physics.  With better data, some clear and independent observational signatures by which to measure particular system parameters may yet be forthcoming -- for example, measurement of BH spin by directly measuring the shape of the photon ring \cite{johnson2020}.  Likewise, sufficiently accurate modeling may reveal reliable trends under certain circumstances -- for example, the $\beta_2$ parameter describing rotational symmetry of the resolved linear polarization, which can reliably distinguish between different magnetic field strengths in simulations, and provides a rough indicator of BH spin if accretion is viewed nearly face-on \cite{palumbo2020}.

However, in general, changes to BH system parameters cause complex and overlapping changes in the observable image and other measurements, which are difficult to disentangle into clear relationships between individual parameters and measured values.  Thus, simulation libraries are still the most reliable way to understand the constraints that new data provide, as well as the most useful current tool available to develop better techniques.  Simulations retain an important advantage over phenomenological models in particular, as predictions can be made about many disparate measurements based on the same underlying simulated model, and thus it can be evaluated against all available measurements of a source in concert.

\subsection{Many Long Simulations}
\label{sec:req_many_long}
A relatively short and sparse parameter study sufficed in order to understand the basic features of \bhname -- however, in order to evaluate the much faster-moving \sgra (\sgr) in detail, longer simulations were required. Ideal GRMHD simulations can be rescaled to the system's characteristic timescale, but the relevant scale is much shorter for \sgr than \bhname, requiring much longer simulations in order to cover just a few days of real time.  Longer simulations also allowed better characterizing the variability of the light-curve, as well as providing more snapshots to compare against the data -- useful considering the less definitive reconstructions of the \sgr image.  This need provided the original motivation for writing \kharma.

Notably, in this case and generally, the advantage to simulation libraries and parameter studies is to produce \textit{many long} simulations, rather than necessarily the \textit{largest} possible single examples.  In GRMHD in particular, so long as basic resolution requirements are met (such as resolving the magnetorotational instability) simulations tend to vary little with resolution -- at least, on the scales at which it is feasible to run an entire simulation library.  Thus a code must produce simulations at the existing resolutions more quickly -- it must use each processor to its full potential (good base performance) and scale efficiently with the addition of more processors to a problem of the same size (good strong scaling).

In particular, the requirement of good base performance also motivates performance portability, as single GPU chips or nodes can vastly outperform single CPU nodes -- thus, the difficult or unavoidable performance penalties of strong scaling can be partially recovered simply because fewer nodes need to communicate in order to achieve the same performance.  There is also great utility in porting simulations which might have required 10-20 CPU nodes onto a single GPU, as single machines are much more likely to be available as a part of campus computing resources, or even within the means of an average research budget.

\subsection{Mesh Refinement}
\label{sec:req_mesh_refinement}
Simulation efficiency benefits from larger time steps, which in turn require the largest possible zones, placing resolution only where it's needed.  This is a universal problem with Logically Cartesian grids in spherical coordinate systems: the allowable simulation timestep set by the Courant-Friedrichs-Lewy (CFL) condition is much smaller than it needs to be, set by the many small zones next to the coordinate pole.  In a single regular mesh, increasing the $\phi$-resolution necessarily arranges more zones next to the coordinate pole, and increasing the $\theta$-resolution causes all zone centers to grow closer to one another as they approach the pole. Both effects reduce the width, and therefore the CFL limit, of these zones, and waste computation time with needlessly small timesteps.

\begin{figure}
    \centering
    \includegraphics[trim={3.2cm 21cm 3cm 3.5cm},width=0.9\textwidth]{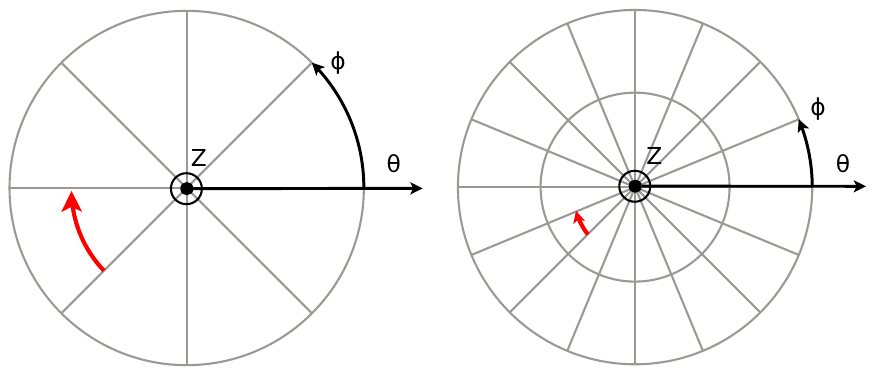}
    \caption{Diagram illustrating the timestep problem faced by non-refined logically Cartesian meshes in spherical polar coordinates.  The illustration represents a slice at constant radius, arranged around the coordinate axis (i.e., $z$-axis) of a spherical polar coordinate grid, as marked (note that exactly the same problem exists in cylindrical or two-dimensional polar coordinates as well, with $\theta$ replaced by the radial coordinate).  As the resolution in the $\theta$ or $\phi$ direction is increased, zones next to the axis bunch together, with smaller zone widths (e.g., those marked in red above). For solvers with a maximum timestep limited to the signal crossing time of the smallest zone (i.e., subject to the Courant-Friedrichs-Lewy condition) this can cause the stability-limited timestep to be much smaller than what would be required to maintain good accuracy, wasting significant computation time.}
    \label{fig:pizza_problem}
\end{figure}

This can be mitigated somewhat by the use of coordinate-based refinement; specially designed coordinate transforms like ``Funky'' Modified Kerr-Schild coordinates \cite{wong2022}, which increase the $\theta$-widths of the circumpolar zones at the coordinate level, so that zones equally spaced in the native coordinate system nevertheless grow in the base spherical coordinate system.  At nominal resolutions, this provides from a 2-8x greater timestep than spacing zones evenly in $\theta$.

However, really addressing this problem requires mesh refinement, which can widen zones in $\phi$ without losing $\theta$-resolution near each coordinate pole.  Successive derefinement of blocks approaching the axis can reduce almost arbitrarily high resolutions in the midplane down to a manageable number of axis zones in the coarsest, polar blocks.  Internal refinement can further reduce the number of zones surrounding the axis to just 4-8 total zones, orders of magnitude fewer zones than an unrefined system, with a proportional speedup.

Other solutions exist for the timestep problem, e.g., overstepping the Courant condition of the circumpolar zones without modifying the mesh geometry \cite{ji2023}.  However, these are complex to implement when compared to mesh refinement, and there is no need for them in a flexible self-contained code.

\subsection{Extensibility}
\label{sec:req_flexibility}
Perhaps the strictest requirement for \kharma short of performance portability was that it be extensible, able to accommodate user changes and new physics onto its fast and scalable platform.  Due to the broad range of physics at play over different scales and accretion rates, greater understanding will require treatment of new phenomena, not just modifications to the problem setup.

It is well known that ideal GRMHD does not capture all the physical effects which operate in most black hole accretion systems, and indeed the most exciting opportunities in really understanding these systems lie in treating physics beyond ideal GRMHD at resolution and accuracy scales where the observational consequences of new physics can be effectively studied.

One target of particular interest was dissipative effects, which may become important at very low accretion rates such as observed in \sgr.  Even the simplest and most streamlined treatment for \sgr \cite{chandra2015} uses a locally semi-implicit scheme as it must treat time-dependent and stiff source terms.  Thus costs from 3-10 times more than fully-explicit ideal GRMHD evolution.  If a more detailed treatment is ever required, it would cost even more.  Resistivity may also play a role at small length scales close to the horizon, which can affect variability and observational appearance -- a resistive treatment requires a similar implicit scheme to dissipative effects, with similar cost.

At the other end of the range of accretion rates, radiative effects quickly become dominant, at first indirectly through radiative cooling, and then directly as radiation pressure.  Radiative effects dictate the evolution of high-accretion-rate systems, which are characterized by bright X-ray emission.  Due to the difficulty of simulating accretion disks with such high luminosity, existing models for these systems are structured around ``thin-disk'' analytic solutions as in e.g., \cite{novikov1973}.  These models have proven tremendously useful, but are limited in their applicability due to the assumptions involved.  Algorithmic treatments of radiative GRMHD are well-understood \cite{ryan2015,white2023}.  However, global treatments are computationally expensive -- computational power is only now reaching the point that full thin-disk radiative simulations at high accretion rate are possible, potentially revolutionizing the field.

As mentioned, algorithms exist for all of these extensions.  However, generally they are expensive, ranging from 3-20 times the base computational cost of ideal GRMHD simulations.  Thus, they have traditionally been simulated at low resolutions, over short timeframes, and only with isolated simulations (not parameter studies).  With a performance-portable framework, implementations of these extra effects can finally be incorporated into libraries of longer simulations, dramatically increasing their utility in comparing to observational data by parameterizing over the system's unknown incidental features.

In short, \kharma's performance portability finally makes feasible a number of exciting simulations treating computationally expensive new physics, by making available the increased computational power of GPUs and the exascale machines which rely on them.  Providing easy interfaces for these new features was therefore a primary goal for the code.

\subsection{Readability}
\label{sec:req_readability}
Finally, any code which expects to see adoption in academia must be understandable to a domain scientist graduate student using the code for a particular project, and not primarily tasked with its development.  Large, complex and multipurpose scientific codebases have a place in laboratory science and engineering in industry, but codes in academia cannot rely on the same level of long-term staffing and funding, and must be less complex to be applied productively.  This is not a strict requirement of simplicity, but rather a dual requirement of modularity and documentation: \kharma cannot assume a high level of institutional knowledge, and must be possible to modify without deep knowledge of its various components.

Many academic codes take a strict simplicity-first approach to this problem, restricting themselves to features and implementations which will be understandable in full to a new student, without requiring that months be devoted to simply learning the code.  It is arguable whether this approach is strictly necessary, and \kharma was designed under the assumption that it is not, as the other desired features and dependencies are simply not compatible with a minimal codebase.  \kharma instead takes a documentation- and standards-based approach to the problem: the hope is that with a proper introduction, new developers need understand only the package interface and a few relevant packages in order to get started.  However, while incredibly successful in most software, physics simulation codes provide something of a worst case for this approach, since debugging numerical instabilities can involve understanding the interactions between different packages modifying the fluid state.  \kharma's long-term success in taking this alternative approach remains to be seen.

\section{Design for Requirements}
\label{sec:design}

To summarize the previous sections, \kharma was required to:
\begin{itemize}
    \item Run simulation steps quickly by leveraging specialized architectures
    \item Simulate the same time periods with fewer steps using mesh refinement
    \item Easily support additional physics in combination with ideal GRMHD, such as viscosity and radiative processes
    \item Still be understandable to the average graduate student not primarily tasked with its development
\end{itemize}

\kharma's major design choices are very explicitly geared toward meeting these four requirements, so the design is presented here in reference to how it helped to address each requirement.

\subsection{Performance Portability}
\label{sec:design_performance_portability}

Performance portability was the core requirement, for reasons exhaustively catalogued in \ref{sec:computing}, and is provided by Kokkos\footnote{\url{https://github.com/kokkos/kokkos}} \cite{carteredwards2014,trott2021,trott2022}.  The Kokkos programming model enables performance portability by providing a vendor-neutral library, compiled alongside an otherwise standard C++ program.  It was chosen as the programming model for Parthenon, and therefore \kharma, due to its simple syntax, universal support, and high performance ceiling relative to native performance.

Kokkos is an effort primarily of Sandia National Laboratory, and supports the major CPU and GPU architectures used in HPC.  In the Kokkos context, ``performance portability'' means enabling the use of a single primary source code base to generate efficient compiled code for many different HPC architectures.  Kokkos is designed to be compiled into C++ programs: calls into the library from user C++ code cause the compiler to transform certain user functions into computational kernels in a number of different languages and language extensions, including \nvidia's CUDA, Khronos SYCL (used for Intel GPUs), and AMD HIP.  The transformation happens entirely at compile-time, without wasting cycles performing just-in-time (JIT) compilation when the program is run. As it supports the use of any C++17-compliant compiler, most vendor programming models, and nearly all accelerator hardware in use in HPC, Kokkos remains almost universally portable yet entirely vendor-neutral.

While this choice was made in 2020, before the universal availability of OpenMP offload implementations, OpenMP offload still does not meet our requirements of near-native performance or broad flexibility.  Besides OpenMP, the competitor technologies were C-like systems such as OpenCL, as well as various ``turnkey'' JIT-compiled solutions.  Modern options such as SYCL and {\tt std::parallel} are also worth considering.  In general, however, the existing options are all designed either for convenience when porting existing code, or for writing specifically GPU-first code.  As a single programming model for CPUs and GPUs, Kokkos abstracts away the divide between CPU and GPU code entirely, enabling performance portability as well as a few convenience features at little overall cost to complexity.

The primary changes to user code under Kokkos are the abstraction of looping and memory access: the user writes parallel code once in an abstract way, and the memory layout and loop ordering are decided at compile time based on the compile target architecture.  In part because of this heavy abstraction, Kokkos provides an easy and natural GPU programming model for the beginner.   Take a normal one-dimensional loop in C:
\lstset{language=C++,basicstyle=\ttfamily}
\begin{lstlisting}
for (int64_t i = 0; i < N; ++i) {
    c[i] = a[i] + b[i];
}
\end{lstlisting}
In Kokkos it is written in a similar way, albeit with very different results:
\begin{lstlisting}
Kokkos::parallel_for(N,
    KOKKOS_LAMBDA (const int& i) {
        c(i) = a(i) + b(i);
    }
);
\end{lstlisting}

The latter snippet replaces the loop declaration with a function call into the Kokkos library, and the loop body with the definition of a C++ lambda function to be run once with each value of the index as its argument.  Use of this construct is indication from the user that there are no loop dependencies, so Kokkos will map loop iterations to the available compute units on the target device in parallel as efficiently as possible, be it CPU, GPU, etc.  The lambda function uses capture by value, so it operates almost as transparently as a normal loop body.

Multi-dimensional arrays (in Kokkos, {\tt View}s) are also accessed in a straightforward way, substituting an inlined function call {\tt ()} for pointer-based array access {\tt []}.  Multi-dimensional and nested loops are a bit more complex, but all of Kokkos's loop formulations are ultimately much more usable than manually specifying thread and block geometry in CUDA, for example.  While the user does eventually have to understand that the host and device do not share a memory space, most host-to-device copies are automatic through the lambda capture, allowing the user to write code nearly identical to what would be written in a host-only {\tt for} loop.  It's important to note this design enables performance portability without obvious GPU-first design overhead such as dictating block/thread breakdowns, and without separating the CPU and GPU codepaths.

In fact, by providing integrated bounds checking in debug builds, automatic memory management tied to the lifetime of the {\tt View} object, and transparent multidimensional array handling, Kokkos can make some aspects of programming and debugging significantly \textit{easier} for the beginner than C, especially when debugging common code errors such as indexing and memory initialization.  After an initial learning curve understanding the syntax, development using the Kokkos abstractions is efficient, and the added features save time.  Code can be developed and debugged locally on the CPU, and only once working compiled for GPU and checked for any (rare) GPU-specific bugs.

It is worth noting a Kokkos-based implementation is not forever locked into Kokkos, either.  The basic lambda-argument interface for kernels is now quite common, shared by many performance portability tools.  With a simple substitution of syntax, the above example could easily be rewritten with C++ {\tt std::parallel}, or using SYCL calls.  As mentioned, Kokkos has a backend for generating SYCL code automatically: it is perhaps indicative that despite having feature parity, this backend is only about half the size of the translation layer for CUDA (4442 non-comment lines vs 7562 as of February 2024).  Especially for \kharma, which mostly uses simple loop kernels, transitioning to e.g. SYCL would be straightforward.

\subsection{Mesh Refinement}
\label{sec:design_amr}

Importantly for \kharma's design, Adaptive Mesh Refinement (AMR) cannot easily be added to a code not designed to accommodate it: keeping track of meshblocks at different refinement levels in a structured-mesh code requires a number of invasive changes to the systems handling coordinates, domain boundary communications, file operations, and (if the mesh layout is to change during the run) also the memory backing and load-balancing.  Thus if AMR is desired, it should be planned into a code from the beginning, if possible.

Parthenon\footnote{\url{https://github.com/lanl/parthenon}} \cite{grete2022} was selected as a dependency partially because it was designed to support AMR -- it makes the construction of an AMR code feasible for a small team by handling nearly all aspects of mesh refinement, allowing the developer to focus on the algorithm as if writing for a single block.  This abstraction breaks down in places, but overall makes the experience of writing and maintaining AMR code much more similar to any traditional multi-block code.

\kharma has carefully preserved AMR support for unmagnetized runs since its inception, regularly testing simple unmagnetized AMR operations to ensure no single-level assumptions crept into the code.  Due to this planning, when a face-centered constrained magnetic field transport was added last year (finally allowing a divergence-preserving prolongation operator, and thus prolongation/restriction of the B field), \kharma could immediately support SMR and AMR meshes for all runs.

In addition to AMR, Parthenon provides tools for nearly all other aspects of an Eulerian structured-mesh code except the physics: parameter handling, file I/O, dynamic tasking, MPI communication, and some basic code organization tools are all provided.

\subsection{Extensibility}
\label{sec:design_adaptability}

Even the fastest code cannot be restricted to performing only ideal GRMHD simulations if it is to be useful going forward.  To be useful, modification of \kharma should be as simple as possible, in order to run new problems, add new features, and simulate new physics beyond ideal GRMHD.

Parthenon provides a pattern for separating code functionality into ``packages.''  Packages do not have strict interface requirements or traditional class inheritance, but rather consist of:
\begin{enumerate}
    \item A C++ namespace for all of a package's unique functions and classes
    \item A {\tt Params} map object (much like a {\tt dictionary} in Python) for storing non-grid-based information such as run-time parameters.
    \item A list of {\tt Fields} (and {\tt Sparse Fields} and {\tt Swarms}) indicating variables which should be allocated on each new {\tt MeshBlock}.  {\tt Fields} can be tied to the size of the {\tt MeshBlock} (i.e., number of cells, faces, edges...), or independent of it.
    \item Some number of ``callbacks'' -- functions implemented by the package which are called at specified points by the code (e.g. a source term, or diagnostics to be printed after every step)
\end{enumerate}

Each package thereby encapsulates code with its relevant data (parameters and fields) similarly to an object in Object-Oriented Programming, with the advantage of translating neatly to templated, device-first implementations.  However, packages usually contain more functionality than individual objects, and are only instantiated once to handle large pieces of program functionality, rather than many times to handle specific instances of data.

To elaborate: in Parthenon, the data itself (particular arrays corresponding to fields) is handled by traditionally object-oriented code in {\tt MeshBlock}, {\tt Mesh}, and {\tt View}.  A package registers fields to be added automatically to each new {\tt MeshBlock}, and then registers callbacks to manipulate that data at particular points during the run (e.g., before a step, when exchanging boundaries, before output, etc.).  This design ensures that packages are neutral to the mesh structure and decisions, but rather represent processes and steps which modify variables on the mesh.

The encapsulation of packages has allowed \kharma to accommodate several different extensions and projects within the same codebase, sharing features upstream in a way that does not impinge on other uses of the code.  Rather than forking for each new project, and eventually implementing overlapping sets of features in several fundamentally incompatible downstream codes, \kharma's package structure encourages centralizing changes, allowing everyone to benefit quickly from improvements.  Two particular elements of the package structure enable this:
\begin{enumerate}
    \item Disabled packages are \textit{disabled}.  Invoking code in a package which is not loaded will nearly always crash rather than producing subtle errors.  Nearly all additional features peculiar to a package can be encapsulated within its own namespace and code, hooking into the main loop with callbacks or behind checks, reducing feature conflicts as much as possible.
    \item Package-based implementations rarely modify the same code and cause conflicts, allowing more developers to work on forks independently for extended times, while still allowing the completed features to be brought back upstream later.
\end{enumerate}

Extensiblity also motivates developing \kharma as a GRMHD-first code, as opposed to a general-relativistic extension to an existing MHD framework.  By incorporating differential geometry from the ground up, \kharma can easily run simulations in any theory with a metric, not just General Relativity, simply by implementing the metric once.  Additionally, it is easy to implement new coordinate transformations, which can be used to compress and expand resolution in a more granular way than block-based mesh refinement.

\subsection{Simplicity}
\label{sec:design_simplicity}

The modular package structure necessarily makes for a more complex piece of software than a traditional C-based GRMHD code.  At several points, these design considerations have required sacrificing strict simplicity in order to introduce features.  This is not to say, however, that simplicity is abandoned as a requirement.  Rather, it is treated as a matter of managing an increased scope and ambition in a way that preserves the goals of a simple code.  Generally, simplicity accomplishes two goals, which remain central constraints in \kharma:
\begin{enumerate}
    \item A new user can learn to modify the code with a minimum of required reading, ideally of documentation and comments rather than source.
    \item Modification of the code is as unlikely as possible to introduce regressions in existing functionality, and any introduced are as easy as possible to find.
\end{enumerate}

These are ultimately not constraints on how big a code is (though both encourage smaller codes, of course).  Rather, they are constraints on how someone interacts with and modifies the code, and thus on its structure and documentation.

\kharma is designed and (especially) documented with these requirements in mind.  As described in Section \ref{sec:design_adaptability}, modular code is the key to writing compatible implementations of complex physics.  A modular structure also attempts to enforce that familiarity be required with as little of the code as possible in order to begin modifying it -- implementations which conform to interfaces can (ideally) be assumed work within those interface parameters, reducing the amount of searching and background information necessary to modify another particular module.  Care must be taken to ensure the maintenance of this design, and of course cross-package bugs are still a distinct possibility, but a modular approach pays off when managing and adding many different features, potentially at the same time.

Modularity also allows easily trimming code which is no longer useful.  Several older optional pieces of \kharma's scheme have been superseded by new components and are now rarely used.  When no longer needed for backward compatibility, these can be easily removed to reduce line count and confusion, without any major disruption to the overall code structure.  Similarly, there is no excuse to hold on to features in the conviction that they might one day be useful: the same package code can simply be reintroduced later, modulo any updates to the package interface.

\section{Code Structure}
\label{sec:structure}

The following sections describe \kharma's structure as a code, including describing the primary algorithm, philosophy, and breakdown into packages.

\subsection{Algorithm}
\label{sec:structure_algorithm}
At its base, \kharma implements the algorithm used in HARM, described in \cite{gammie2003}, though it evolves an amended version of the 3-velocity as described in \cite{mckinney2004}.  For primitive variable inversion, it has traditionally used the $1D_w$ scheme from \cite{noble2006}, with the ``Flux-CT'' divergence-preserving magnetic field transport described in \cite{toth2000}.

However, alternatives to nearly every part of the algorithm are available as different packages, which can be enabled to construct a particular algorithm, or emulate another code's choices for comparisons.

For example, four transport schemes are available for the magnetic field: in addition to ``Flux-CT,'' \kharma implements the face-centered constrained transport scheme outlined in \cite{stone2009}, including a number of schemes for determining the EMFs along zone edges with and without upwinding.  Also, though these are rarely used now, it implements a constraint-damping scheme similar to \cite{dedner2002} usable in flat spacetimes, as well as a projection operator approach (described in \cite{toth2000}) using a stabilized Biconjugate Gradient (BiCGStab) solver \cite{vandervorst1992}.  These schemes are separated into different packages and made mutually exclusive at run-time so as not to interfere with one another.

In addition to the $1D_w$ primitive variable recovery noted above, \kharma now implements the robust primitive variable recovery of \cite{kastaun2020}, which greatly improves stability.

\kharma implements a first-order flux correction (FOFC) scheme \cite{beckwith2011}, which speculatively determines which zones will encounter floors or inversion failures, and pre-emptively uses first order, Local Lax-Friedrichs fluxes with Donor Cell reconstruction to update just those potentially problematic cells, providing extra stability and less injection of floor material, at the cost of using a less accurate scheme in just those zones which would have required intervention.  Combined with reliable variable inversion, this allows significantly more stable performance when integrating shocks at high magnetization, allowing for the introduction of fewer stability floors in the simulation.

\kharma also implements a large number of different variable reconstruction schemes (WENO5 \cite{liu1994}, WENOZ \cite{castro2011} including a linearized robust form\footnote{To be characterized in Barker et al (2024)}, MP5 \cite{suresh1997}, PPM \cite{colella1984}, PPMX \cite{colella2008}, linear with Van Leer and Monotonized Central limiters \cite{vanleer1974,vanleer1977}, and Donor Cell \cite{godunov1959}).  Other reconstruction schemes can be added easily with just a single-zone implementation in C or C++.

For time integration, \kharma supports not only the traditional predictor-corrector update used in most HARM codes, but RK1, RK2, and SSPRK(5)4 schemes as well -- the \kharma sub-step is implemented so as to be independent of the integrator used, so any integrator implemented in Parthenon can be immediately used in \kharma.

When the density or internal energy fall below predetermined values, or become too small relative to the magnetic pressure, extra material and energy are directly injected to keep stability.  \kharma supports injecting floor material in any of a number of different frames, the most common of which is the drift frame (as tested in \cite{ressler2017}).

\kharma also supports theories with arbitrary implicit source terms, such as the ``Extended GRMHD'' theory of \cite{chandra2015} treating viscosity and heat conduction in Coulomb-collisionless plasmas (see Section \ref{sec:feature_solver} for details).  When evolving a semi-implicit theory, \kharma supports only a predictor-corrector scheme for integration, as it must be augmented by a nonlinear solve for the new state after each sub-step.  In addition, the implicit solver doubles as the primitive variable recovery, so no traditional primitive recovery solver is used.

\subsection{Package structure}
\label{sec:structure_packages}

\kharma relies on Parthenon as its single direct dependency, and this dictates much of its overall code structure and conventions.  Parthenon provides interfaces to make use of MPI communication, HDF5 parallel file operations, and Kokkos loops and {\tt View}s -- \kharma then uses these features almost exclusively through Parthenon.  Parthenon additionally provides a framework for code modularization, ``packages,'' which \kharma uses heavily.  In fact, perhaps the best introduction to the code structure is to look at the internal dependency structure between \kharma's many packages.

\begin{figure}
    \centering
    \includegraphics[width=\textwidth]{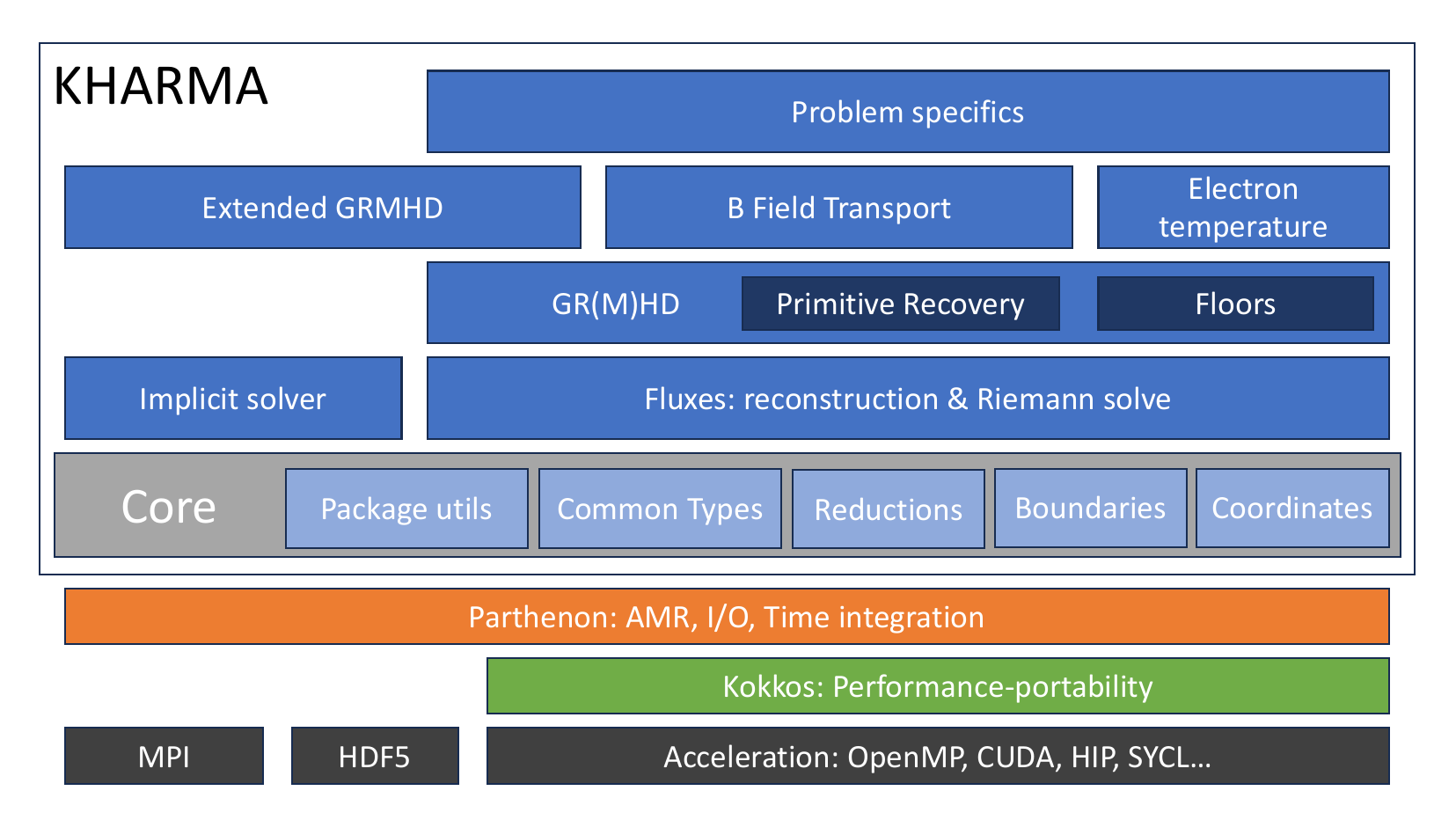}
    \caption{Diagram of \kharma's structure and dependencies.  Dark grey boxes at bottom are system dependencies, above those are the bundled dependencies Kokkos and Parthenon.  The large outline marks \kharma proper. The lightest blue boxes within the grey outline denote core \kharma features on which all packages depend. Darker blue boxes are \kharma packages, with the darkest boxes indicating fully-interdependent ``sub-packages.''  Placement of a module above another module or group denotes a direct dependency.}
    \label{fig:kharma_dia}
\end{figure}

A full diagram of \kharma's package structure is provided in Figure \ref{fig:kharma_dia}.  Starting from the bottom row, it shows system dependencies of \kharma: MPI, HDF5, and an accelerator framework: usually OpenMP, CUDA, HIP, or SYCL.  (\kharma obviously also requires a C++17-compliant compiler.)  Next are the dependencies compiled alongside \kharma: Kokkos and Parthenon.  Finally, items within the box are \kharma components, almost universally packages.

\kharma's components are effectively split into four levels, detailed in the following four sub-sections.

\subsubsection{Core}
\label{sec:packages_core}

The core level consists of mostly convenience features, parameters, and common datatypes -- anything common to the whole code, which could be used in any other package.  Among these only {\tt Globals}, {\tt Reductions}, and {\tt Boundaries} are formal packages.

The {\tt Package} core module/namespace defines the standard interface for packages in \kharma, which adds features to the Parthenon definition for convenience.  \kharma provides several new callbacks for its packages, structured around common points in a GRMHD update step; e.g., for adding source terms, operator-split physics, or conversions between primitive and conserved variables.

Other core modules house common data structures and code for easily calculating index ranges, as well as code for performing reductions among MPI ranks, used for diagnostics and summary statistics by several different packages.  The core modules also include a {\tt Globals} package, which provides more universal access to a couple of useful variables, such as the current simulation time and state (initializing vs. stepping).

Finally, the core includes a differential-geometry-aware coordinates object, {\tt GRCoordinates}, which takes the place of Parthenon's built-in coordinates. A new {\tt GRCoordinates} object is created automatically with each {\tt MeshBlock}, providing data for locations within that block.  In addition to cell locations in the code's coordinates, as provided by Parthenon, {\tt GRCoordinates} introduces the concept of a coordinate transformation between the code's coordinates (in which zones are evenly spaced) and a system of ``embedding'' coordinates representing the basic coordinates for the spacetime (e.g. Spherical Kerr-Schild).  The {\tt GRCoordinates} object also carries caches of expensive-to-calculate quantities such as the metric and connection coefficients.

\subsubsection{Solvers}
\label{sec:packages_solvers}
The next level of abstraction implements reconstruction schemes, Riemann solvers (in the {\tt Flux} package), and semi-implicit stepping (in the {\tt Implicit} package).  Together, these are intended to form a framework for solving any hyperbolic system in curved spacetime via finite volumes, allowing for the same code to be shared between different schemes with or without magnetic fields, viscous effects, radiative feedback, etc.

This separation is imperfect still, as the Riemann solver requires determining sound speeds and states of conserved variables from primitive variables, both of which are theory-dependent but must live in the same kernels as the Riemann solve itself, to maintain performance.  However, the distinction is enough that \kharma supports a few theories (hydrodynamics, magnetohydrodynamics, and extended MHD) in a relatively natural way with a minimum of code duplication.

The {\tt Flux} package also handles correcting any fluxes which will lead to floor hits or primitive recovery failures (i.e., first-order flux corrections, see Section \ref{sec:structure_algorithm}).

The implicit solver is written to be general, allowing any variables marked with the tag {\tt Metadata::Implicit}, to be evolved semi-implicitly.  Any time-dependent source terms acting on the new variable must still be added manually, however.  The implicit solver is described in Section \ref{sec:feature_solver}.

\subsubsection{Physics}
\label{sec:packages_physics}
The {\tt GRMHD} package handles updating the primitive hydrodynamic fluid variables associated with a HARM scheme -- density, internal energy density, and velocity.  It does not evolve the magnetic field components, however.  Thus, while it optionally uses a magnetic field to compute conserved variables like stress-energy tensor (and the signal speeds, etc), it is agnostic to the magnetic field transport scheme, and supports several different transports implemented in different packages.

When evolving the GRMHD variables explicitly, the {\tt Inverter} package handles the numerical inversion of conserved to primitive fluid quantities (when evolved implicitly, the variable inversion is done automatically by the nonlinear solver).  As they share substantial structure and code, this package implements both of the available primitive variable recovery schemes (see Section \ref{sec:structure_algorithm}), which can be selected at runtime.

The {\tt Floors} package, implementing various density and internal energy floors and injection frames, as well as a number of variable limits, also lives alongside/within the GRMHD algorithm.

\subsubsection{Extras}
\label{sec:packages_extras}
There are a few additional packages which enhance the main GRMHD package.  Generally, these implement any additional physics beyond evolving the core GRHD variables: electron temperatures, dissipative effects, and so on.

These packages also include the magnetic field transports, solutions for evolving the magnetic field while preserving low divergence.  \kharma supports several different magnetic field transport schemes, but modern simulations generally use the face-centered constrained transport scheme unless another package is required for backward compatibility, testing, or speed.

\subsubsection{Glue}
\label{sec:packages_glue}
Time-stepping is orchestrated by the {\tt Driver}, which constructs a list of tasks constituting a single simulation sub-step.  \kharma implements three different time-steps, which produce near-identical results but differ in which variables they synchronize, and the order they update: the first type mirrors most HARM codes in treating the primitive variable state as the fundamental representation, exchanging primitive variable values at boundaries.  The second treats conserved state as fundamental and exchanges conserved variables, allowing consistent evolution with just one synchronization per sub-step in some circumstances (generally, if primitive variable recovery failures may need to be fixed, HARM requires two synchronizations).  The last step type is a simplified algorithm for debugging and testing, without the additional tasks needed for AMR or face-centered mangentic fields.

Fluid initialization code can also take advantage of any package: different problems can require or make optional whatever fields make sense in context.  Where there is a common, standard initialization which makes sense (i.e. a nominal ``cold'' value for electron temperatures which will be heated to equilibrium by the fluid) this is applied if the package is loaded but the field is not otherwise initialized.  Initialization routines which require fields not enabled by the user generate errors.

Since different magnetic field transports preserve different representations of the magnetic field divergence, initialization of the magnetic field is dependent on what field transport will be used, and thus is handled separately from problem initialization.  Where problems require a specific magnetic field configuration, they can set parameters for this later initialization, so as not to have to implement two magnetic field initializations per problem.

While it contains quite a few different modules, \kharma is not (yet) a terribly complicated code overall -- Figure \ref{fig:kharma_dia} lays out almost all of the code modules and internal dependencies.  Freed from implementing the extensive bookkeeping of mesh refinement or specialized syntax of device code, \kharma consists almost exclusively of physics-related code, which makes up 3 of the 4 main layers.

\section{Code Features}
\label{sec:features}

\subsection{Implicit Solver and Extended Theory}
\label{sec:feature_solver}

\kharma allows for semi-implicit time-stepping in limited instances, using a non-linear solver to obtain field primitive values at the end of a substep after any conserved fluxes have been applied explicitly.  This operation is applied separately to each zone, and necessarily after the fluxes have been applied, allowing \kharma to accommodate theories with implicit source terms.

The solver itself is a multi-dimensional Newton-Raphson root finder that minimizes a vector-valued loss function over the primitive variables, representing their state at the end of the timestep.  From an initial guess consisting of primitive variables at the beginning of the timestep, it computes a Jacobian via the secant method, and evaluates a correction via linar matrix solve (we use a pivoted version of the serial QR decomposition from the {\tt kokkos-kernels}).  The process is iterated to desired accuracy, optionally with a backtracking line search for robustness.

The package is able to evolve an arbitrary subset of the overall variables implicitly, from a single-variable solve as in moment closure or constraint-damping schemes, to the semi-implicit evolution of all fluid variables. The latter is required for the ``Extended GRMHD'' scheme of \cite{chandra2015}, which \kharma fully supports.

\subsubsection{Extended GRMHD}
\label{sec:egrmhd}

The Extended GRMHD (EGRMHD) theory adapts the equations of ideal GRMHD to include shear viscosity and heat conduction, in a simple way suitable for Coulomb-collisionless systems such as the accretion of material onto \sgra. In such systems, the length scale for Coulomb collisions $\lambda_{\mathrm{mfp}}$ is much longer than the dynamical length scale $r_g$, which is longer than the Larmor radius $\rho_c$.  This implies collisionality perpendicular to the magnetic field, but collisionless kinetic behavior parallel to it.  This separation of directions forms the main assumption underlying the simplifications in EGRMHD.

EGRMHD treats only two additional variables, beyond the eight in ideal GRMHD.  The first is $\Delta P$, the pressure anisotropy with respect to the local magnetic field (since isotropy will be maintained in directions perpendicular to the field).  The second is $q$, the scalar heat flux along magnetic field lines.

\begin{figure}
    \centering
    \includegraphics[trim=0 0.2cm 0 0.2cm,clip,width=0.48\textwidth]{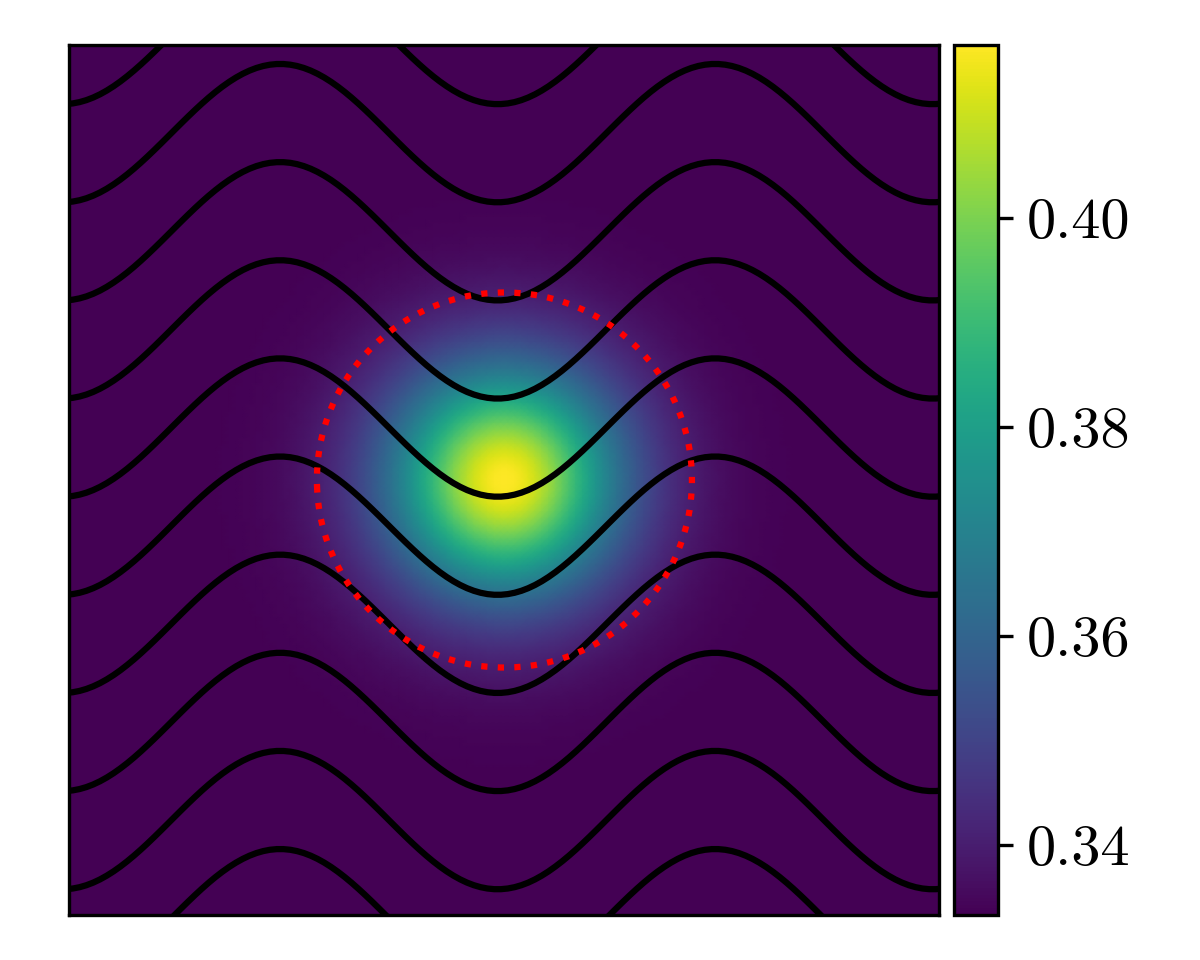}
    \includegraphics[trim=0 0.45cm 0 0.4cm,clip,width=0.51\textwidth]{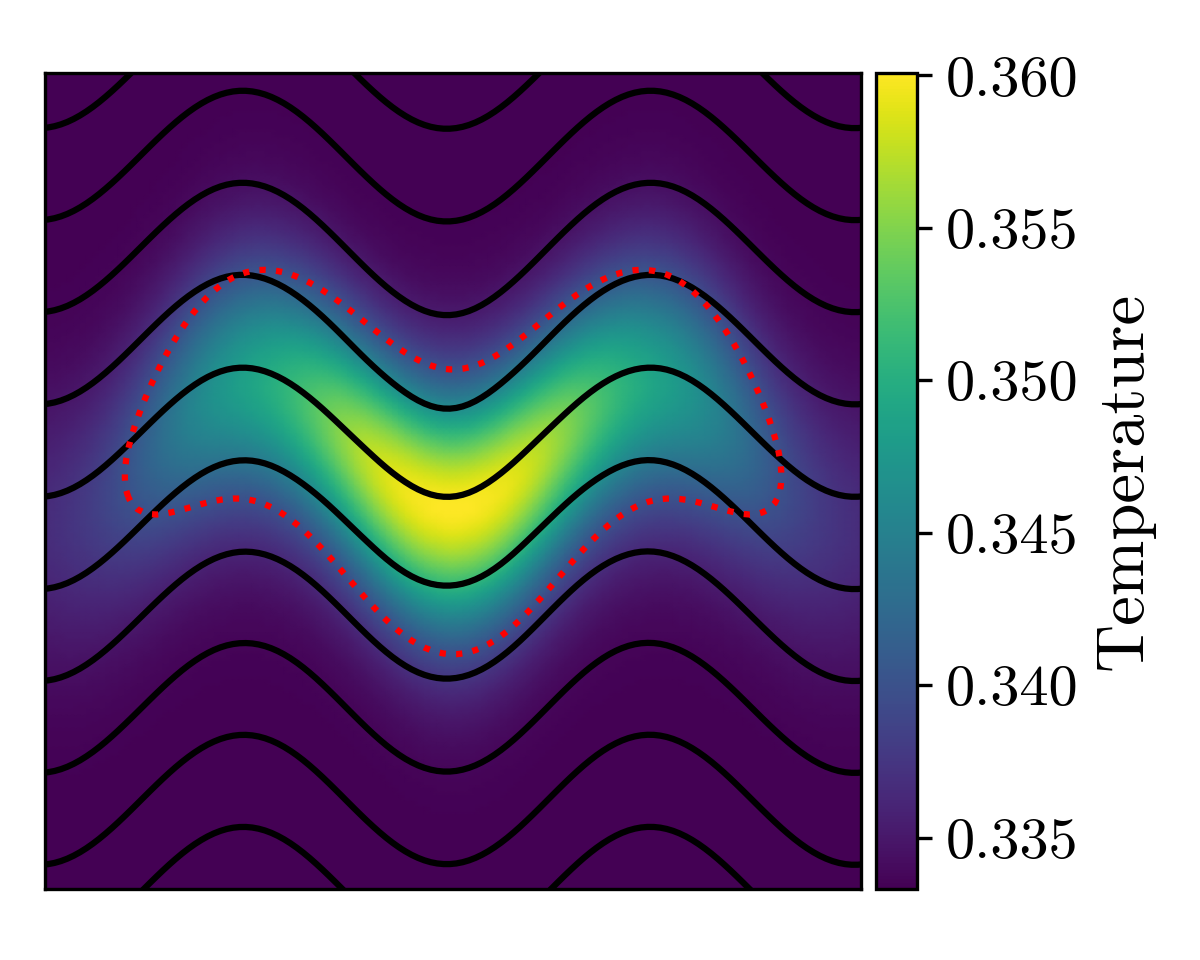}
    \caption{Anisotropic conduction of heat in strong magnetic field, simulated in \kharma by solving the equations of Extended GRMHD using a semi-implicit scheme.  At left is a plot of the temperature in code units at initialization, with a single circular hotspot initialized in the center.  The dotted red contour denotes the extent of non-background temperature, i.e. $> 3\%$ above the background value.  At right is a later snapshot, showing the thermal energy diffused via conduction along the field lines.}
    \label{fig:anisotropic_conduction}
\end{figure}

Figure \ref{fig:anisotropic_conduction} illustrates the scheme by modeling the anisotropic heat conduction on an artificial background magnetic field.

We provide a brief description of the EGRMHD equations following \cite{foucart2016}.  In contrast to the ideal GRMHD stress-energy tensor, the now-familiar
\begin{align}
T^{\mu\nu}_{\rm ideal} = (\rho + u + P + b^2)u^\mu u^\nu + (P + b^2/2) g^{\mu\nu} - b^\mu b^\nu
\end{align}
EGRMHD evolves a version with additional contributions,
\begin{align}
T^{\mu \nu} = T^{\mu\nu}_{\rm ideal} + q^\mu u^\nu + q^\nu u^\mu + \Pi^{\mu\nu}
\end{align}
where $q^\mu = q \hat{b}^\mu$ is the heat flux, $\hat{b}^\mu = b^\mu/b$ is the unit vector along $b^\mu$,
\begin{align}
\Pi^{\mu\nu} = - \Delta P \left( \hat{b}^\mu \hat{b}^\nu - \frac{1}{3} h^{\mu\nu} \right)
\end{align}
is the viscous stress-energy tensor, and 
\begin{align}
h^{\mu\nu}=g^{\mu\nu}+u^\mu u^\nu
\end{align}
is the projection on a hypersurface orthogonal to the 4-velocity $u^\mu$.  The total stress-energy tensor is evolved according to the Bianchi identity as usual, and the evolution equations for density and magnetic field are unchanged from ideal GRMHD.  This leaves the evolution equations for the heat flux and pressure anisotropy. 
These are written with respect to rescaled variables
\begin{align}
\tilde{q} =& q \left(\frac{\tau_R}{\chi \rho \Theta^2}\right)^{1/2},\label{eq:q}\\
\Delta \tilde{P} =& \Delta P \left(\frac{\tau_R}{\nu \rho \Theta}\right)^{1/2}\label{eq:dP},
\end{align}
as
\begin{align}
\nabla_\mu (\tilde{q} u^\mu) =& - \frac{\tilde{q}-\tilde{q}_0}{\tau_R} + \frac{\tilde{q}}{2} \nabla_\mu u^\mu, \label{eqn:qevol}\\
\nabla_\mu (\Delta \tilde{P} u^\mu) =& -\frac{\Delta \tilde{P} - \Delta \tilde{P}_0}{\tau_R} + 
\frac{\Delta\tilde{P}}{2} \nabla_\mu u^\mu. \label{eqn:dpevol}
\end{align}

That is, $\tilde{q}$ and $\Delta \tilde{P}$ are damped towards their target value $\tilde{q}_0$ and $\Delta\tilde{P}_0$ over a timescale $\tau_R$. The evolution equations follow from the requirement that the model satisfies the second law of thermodynamics, as detailed in \cite{chandra2015}, and reflect the change of variable introduced in \cite{foucart2016}.  Both papers additionally detail the closure model used for setting $\tau_R$, which we will not describe in detail here.

In the context of implementation, it suffices for us to note that due to the covariant derivatives on the right-hand side of (\ref{eqn:qevol}) and (\ref{eqn:dpevol}), the evolution of $q$ and $\Delta P$ must be done locally implicitly, by solving a system of equations involving the time-dependent source terms in order to obtain the state at the end of each step.  This is done using \kharma's implicit solver, by marking the GRMHD and EGRMHD variables (though not the magnetic field!) as implicitly evolved, adding them to the solver's update step.

EGRMHD is a promising avenue for future research, especially in understanding the time variability of sparse accretion systems like \sgra.  Further details on \kharma's implementation, e.g. simulation floors and stability techniques, will be provided along with the first physics results from \kharma simulations using the scheme.

\subsection{Simulation Re-gridding}
\label{sec:feature_regrid}

One of the greatest time investments in running high-resolution simulations is the evolution from initial conditions to the steady state.  With modern simulations using physically-larger disk sizes which take longer to equilibrate, coupled with greater accuracy requirements necessitating an accurate idea of the variability after erasing initial conditions, just erasing the initial conditions can amount to millions of steps of wasted evolution, costing many GPU hours and days of real time.

It is much more efficient to restart a simulation which was initialized at low resolution, and simply interpolate its state onto a finer grid as desired.  \kharma can linearly interpolate between arbitrary grids when restarting from an existing state, allowing simulations to be arbitrarily refined or de-refined when they are continued.  This is automatic -- restarting with files of the correct format, if the specified grid is a different size from the snapshot's grid, \kharma will automatically interpolate to the new grid size and shape.

Linear interpolation of the magnetic field, however, does not preserve the divergence.  Thus, \kharma also automatically cleans the magnetic field of divergence when interpolating, by finding the projection operator to the closest divergence-free field to the interpolated state.  \kharma includes a stabilized biconjugate gradient solver for eliminating magnetic field divergence; this is applied to produce an almost arbitrarily small field divergence, which is then preserved in the subsequent run by any of \kharma's magnetic field transports.

\subsection{Multi-Zone Operation}
\label{sec:feature_multizone}

One pressing problem in understanding large-scale galactic and cosmic evolution is that of supermassive black hole (SMBH) feedback.  Large-scale galactic or cosmic simulations cannot capture individual SMBHs, but must incorporate their accretion and feedback for accuracy. Thus when simulating, they are modeled with simple sub-grid accretion models which make many unverified assumptions.

Ongoing work with \kharma \cite{cho2023} aims to bridge these scales, conducting chained GRMHD simulations from the event horizon scale all the way out to the galactic scales $10^7$ times greater.  In these ``multi-zone'' simulations, the total domain is split into 8 annuli, each spanning two orders of magnitude and overlapping by half.  Simulations are conducted of each annulus for a portion of the characteristic timescale, and the fluid state of the simulation's inner portion becomes the outer portion of the next simulation, which is continued for its own characteristic time, and so on.  The resulting chain of simulations is then run repeatedly, allowing material and magnetic field to inflow and escape, eventually reaching a steady-state solution.

Using multi-zone simulations, we hope to dramatically improve the feedback models used in large-scale simulations of galactic and cosmic development, both by producing better semi-analytic characterizations usable in galactic simulations, but eventually also by self-consistently alternating between galactic and SMBH-scale simulations, treating the feedback in full 3D.

\subsection{Run-Time Options}
\label{sec:feature_runtime}

All of \kharma's features, including all problem setups and the selection of any spacetime geometry, are available at run-time.  This means that once compiled, the single resulting binary is capable of doing anything implemented in the code.  Those compile-time flags which do exist are almost purely for changing performance characteristics -- splitting loops for faster operation on GPUs, or using constants for faster operation in flat spacetimes.

Since compile-time options are so limited, they need only be modified when porting \kharma to a new machine, not when running a new problem, test, etc.  Once a particular compiler stack works, it can be codified as a ``machine script'' with the module environment, compile options, etc.  Further compilations of the code can then use a simple invocation of the top-level script {\tt make.sh}, with just a few optional arguments for specifying the type of build (e.g.,~debugging vs production).

One other advantage to run-time options is to allow binary and container distribution.  This doesn't often matter for \kharma yet, as most users are also developers, and generally require builds for GPUs and specialized hardware.  However, containerized \kharma has proven useful a few times already, and may prove more useful in the future as container tools become more universal.

\section{Readiness and Testing}
\label{sec:testing}

At its core, \kharma is an implementation of the well-understood HARM scheme for GRMHD.  It was designed first to closely match its predecessor code \iharm, which was validated as a part of the EHT collaboration GRMHD code comparison \cite{porth2019}.  Where the scheme contains new elements, such as more modern magnetic field transport and primitive variable recovery schemes, these are tested to be drop-in replacements for existing functionality in existing problems.  A particular boon for debugging is that \kharma has enough scheme options that it can usually closely emulate the schemes of other GRMHD codes, to cut down on variables when understanding differences between codes, or specific quirks of behavior.

\begin{figure}
    \centering
    \includegraphics[width=0.32\textwidth]{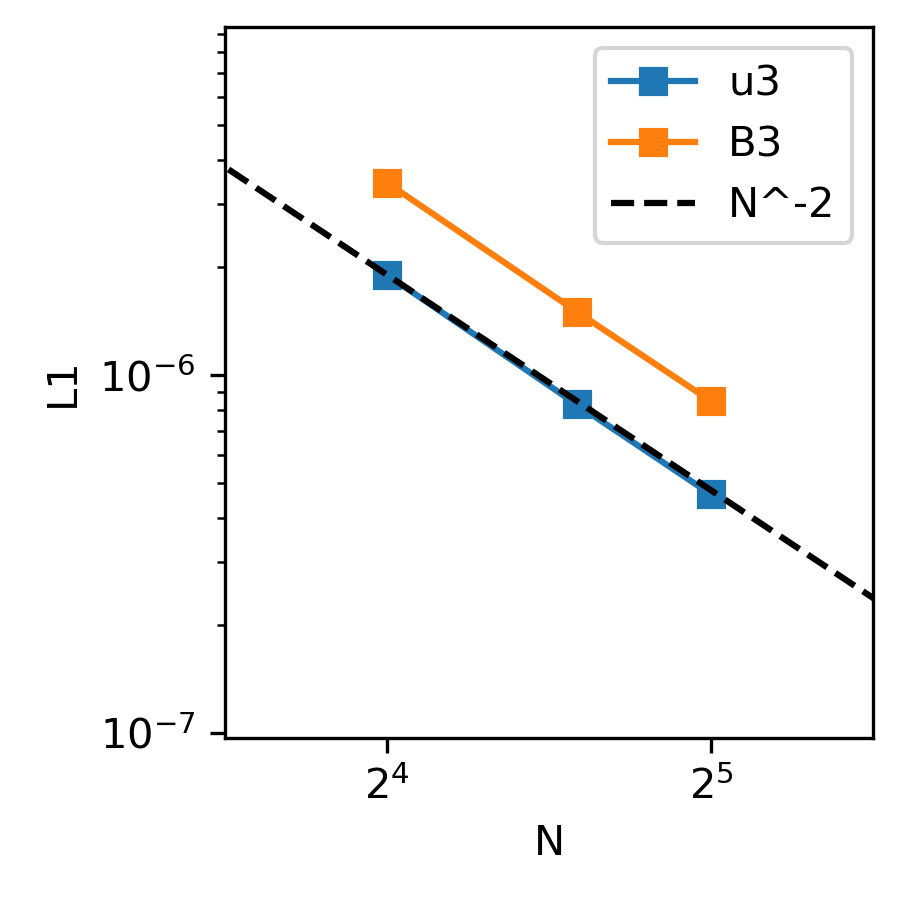}
    \includegraphics[width=0.32\textwidth]{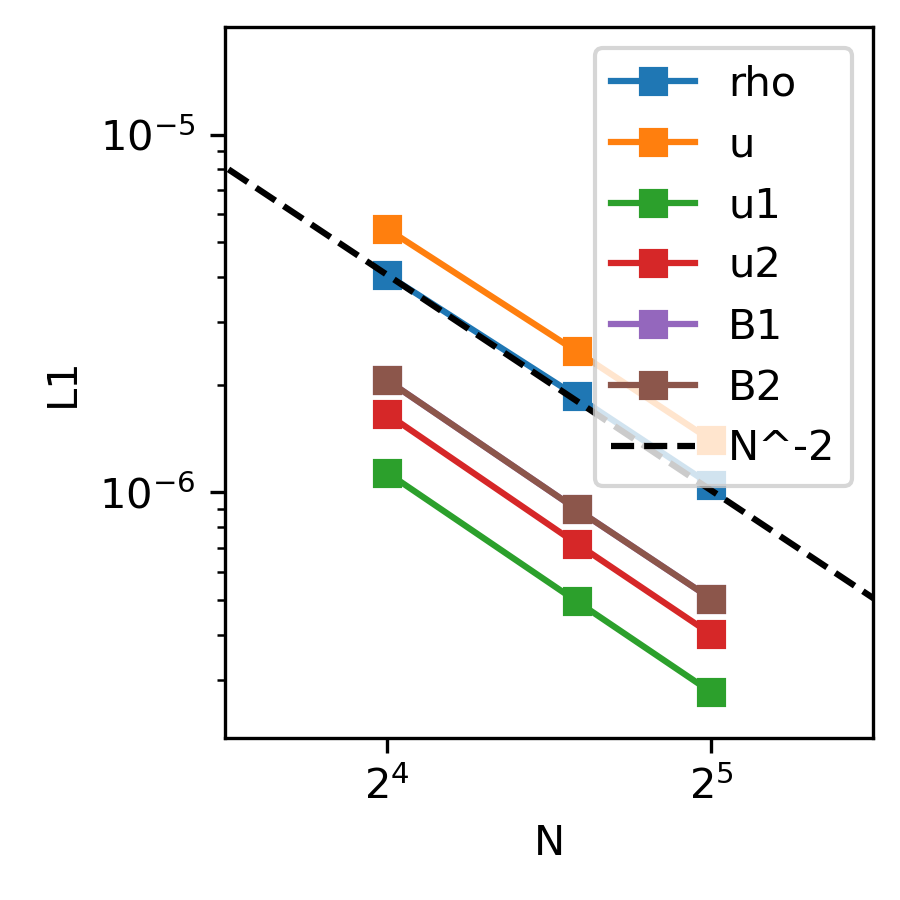}
    \includegraphics[width=0.32\textwidth]{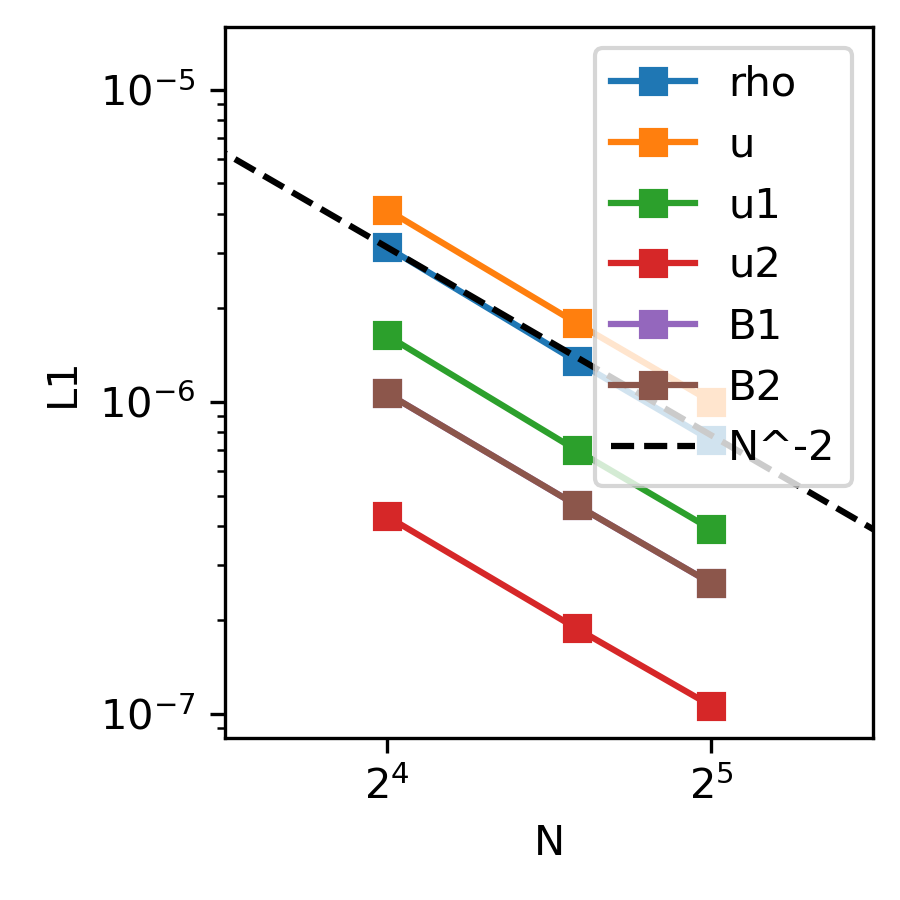}
    \includegraphics[width=0.32\textwidth]{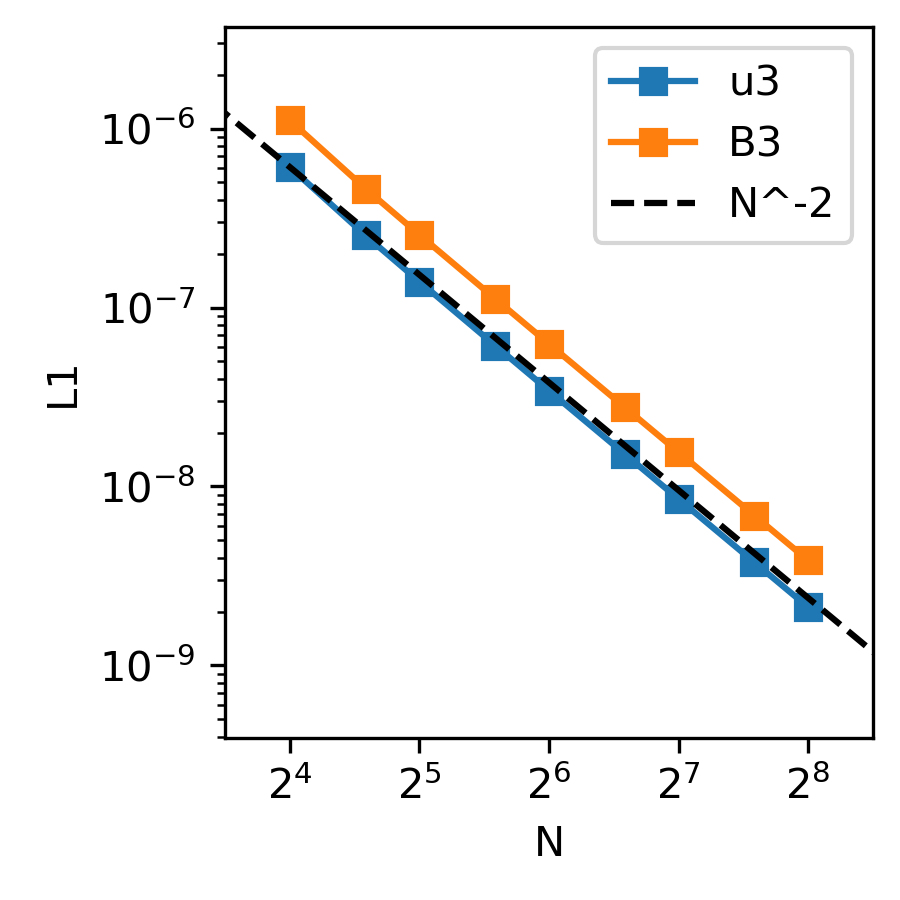}
    \includegraphics[width=0.32\textwidth]{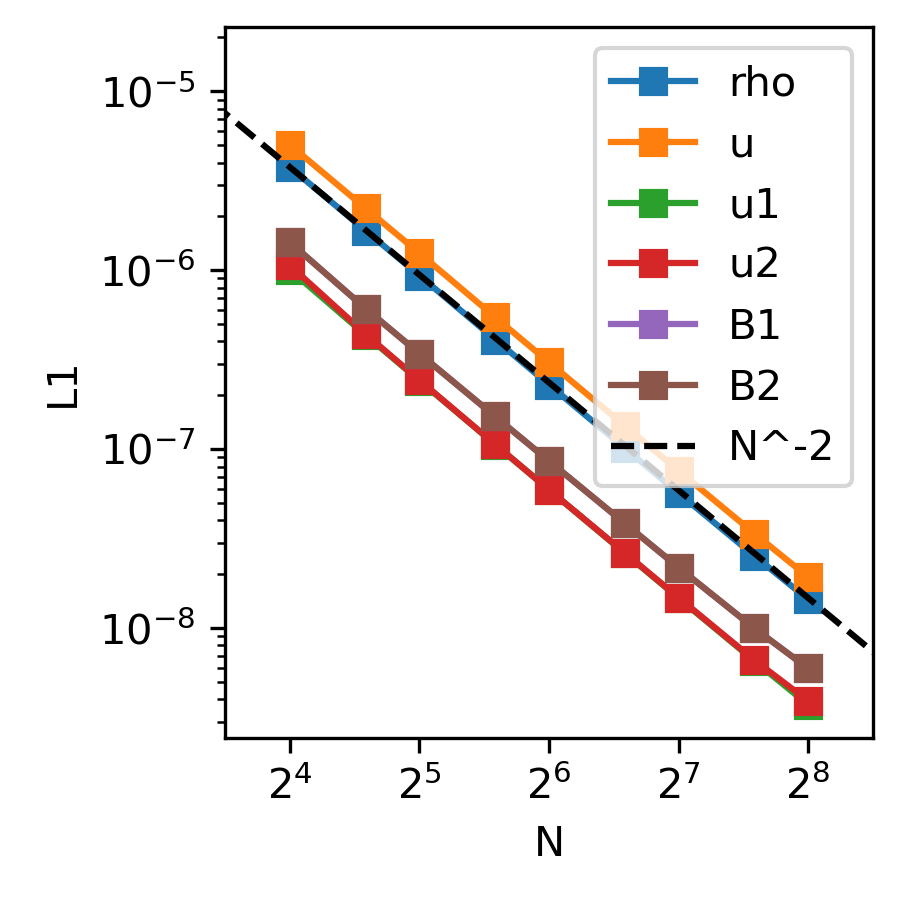}
    \includegraphics[width=0.32\textwidth]{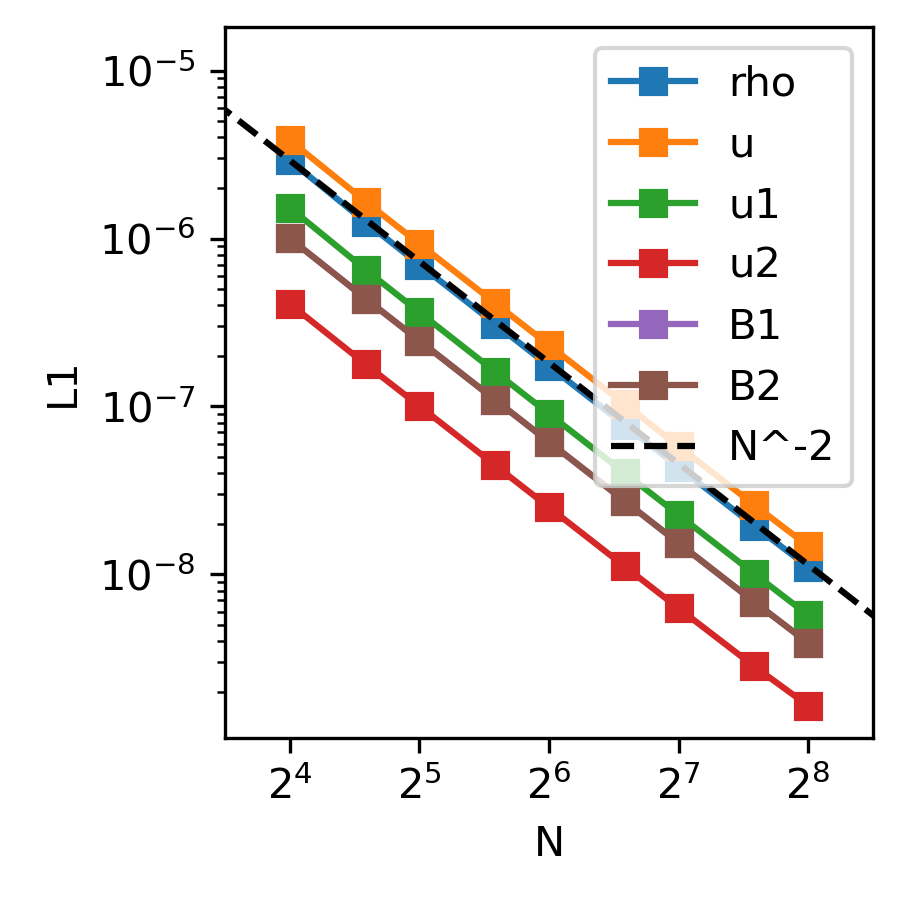}
    \caption{Convergence figures from some of the tests performed on \kharma after every code commit, illustrating convergence when propagating (from left to right) the Alfv\'en, fast, and slow modes in 3D (top) and 2D (bottom).  Each panel plots the L1 norm of each primitive variable when compared to the analytic solution, versus the simulation resolution.  In a second-order scheme, the L1 norm should decrease as the inverse square of resolution, as demonstrated in each test.
    These plots are taken directly from the most recent automated tests at time of writing (hence the independent legends and axes); these and nearly a hundred other figures corresponding to every test configuration are generated automatically on each code commit and available publicly.  A total of 46 linear modes tests are performed, covering all \kharma's available sub-step types,  integrators, magnetic field transports, mesh refinements, and so on.  In addition to convergence on the linear modes problems, another 14 different types of correctness and stability tests are performed -- where applicable, these each cover a wide array of \kharma's different configurations.}
    \label{fig:kharma_convergence}
\end{figure}

Each time new code is committed to the main \kharma repository, a number of regression tests are run: convergence on GRMHD and Extended GRMHD problems (three linear modes and one Bondi accretion problem each) is tested using as many different schemes and features as possible -- at current count, 46 ideal MHD linear modes convergence tests are performed, not counting Extended MHD modes or mesh-refined versions of the problem.  Figure \ref{fig:kharma_convergence} shows these.  The L1 norm is checked to match second-order convergence very closely, usually to scale with resolution at a power between $-1.9$ and $-2.1$.  While \kharma does not have separate unit tests, nearly all relevant codepaths for physics are tested against alternatives as a part of the regression tests, which isolates bugs in most subsystems.  Regression tests are currently run on CPUs automatically, and on GPUs upon manual invocation.

In addition to convergence tests, checks are also performed for determinism upon multiple invocations, and for consistency after restarting from a checkpoint file.  The initialization and (short) evolution of a torus problem is also tested, for error and basic consistency metrics such as the magnetic field divergence.  Both test types include versions with and without mesh refinement.  Another test is run verifying that none of the default parameter files shipped with the code generate errors (due to a change in parameter names or defaults, for example).

Several other well-known standard test problems are implemented which may be run manually, such as the Orszag-Tang vortex \cite{orszag1979}, the magnetized explosion problem of \cite{komissarov1999}, and others.  These are useful for gauging relative performance and stability of different algorithmic choices available in \kharma, or characterizing \kharma's performance against other MHD codes.

Finally, \kharma includes several tests of the electron temperature tracking scheme, adapted from tests of the original implementation in \cite{ressler2015}, and for the extended GRMHD implementation (see Section \ref{sec:feature_solver}), adapted from \cite{chandra2017}.  These will be detailed more closely in a future work, when physics results from these modules are first published.

\section{Performance}
\label{sec:performance}

Scaling tests of \kharma have been performed on many different machines: Stampede2, Frontera, and the Frontera GPU subsystem, Longhorn, at the Texas Advanced Computing Center (TACC), Summit and Frontier at the Oak Ridge Leadership Computing Facility (OLCF), Delta at the National Center for Supercomputing Applications (NCSA) at the University of Illinois, and ThetaGPU at the Argonne Leadership Computing Facility (ALCF).  Results here reflect performance running a torus accretion problem in general relativity using ideal GRMHD, in spherical coordinates, with the $1D_w$ inverter and Flux-CT magnetic field transport, for consistency between different \kharma versions.

\kharma was initially designed and tested for the \nvidia V100 GPU and IBM AC922 architecture used by Summit and Longhorn, where it achieves about 21 million zones by a cycle (ZCPS) per V100 GPU.  This weakly scaled near-perfectly to arbitrary problem sizes/node counts, but more importantly, strongly scaled to problem sizes of interest (e.g.,~$256^3$ total zones) across enough nodes to produce efficient turnaround times of a few days for runs of $10,000\;t_g$.  Numbers from these machines necessarily reflect earlier versions of the code, as \kharma has not been run on these machines recently (indeed, Longhorn has been retired).  Due to a number of new features in the modern code, \kharma versions from this timeframe are generally slightly faster than the most recent commits.

More recently, \kharma has primarily targeted A100 GPUs of Chicoma and Delta, which use the now-common architecture pairing AMD EPYC CPUs with \nvidia A100 GPUs (also used by e.g., Polaris at ALCF and other machines).  Base performance with A100s is promising: running production problems of $128^3$ is nearly twice as fast as on V100 GPUs, or about $35$ million Zone-cycles per second per card.  Thus on one node of Detla (4 GPUs) \kharma achieves the performance of more than 50 nodes of Stampede2.  \kharma has not yet been tested at scale on the new H100 GPUs, which promise some potentially relevant HPC-targeted features as well as higher memory bandwidth than Ampere-generation GPUs.

The current target for performance tuning of \kharma is the AMD MI250X GPUs of Frontier, at Oak Ridge Leadership Computing Facility (OLCF).  Due to register pressure constraints of its complex compute kernels, \kharma achieves only about 100M ZCPS per node of Frontier -- about 25M ZCPS per MI250X card, or 12.5M ZCPS per GPU die.  This is only similar to performance on a node of Summit, despite the increased memory bandwidth and compute available on Frontier -- thus if \kharma were fully memory-bandwidth constrained, performance on Frontier could theoretically improve by a factor of 2.

Finally, it is important to note that performance of \kharma is widely variable, depending on the individual components used.  Different primitive variable solvers, magnetic field transports, spacetime geometries, and stability strategies (floors, corrections, etc.), and the presence of SMR/AMR will significantly affect performance.  The configuration used for these tests is production-ready but relatively simple compared to \kharma's modern feature set, as the hope is to be as directly comparable as possible across many different performance tests over years of development.  More advanced features can run 10-30\% more slowly than the base configuration here.

One feature bears special mention: the semi-implicit mode for running ``Extended GRMHD'' problems (see Section \ref{sec:feature_solver}) is very costly.  On CPUs, \kharma runs EGRMHD problems about five times more slowly than ideal problems, and on GPUs the factor is even worse, between 10-20 times depending on the hardware.  The semi-implicit evolution of EGRMHD problems is more computationally complex than explicit evolution, but performance on GPUs appears to be a factor of 2-4x below what might be expected given CPU performance.  This is because the complex GPU kernels of the nonlinear solver require many registers for each thread, severely reducing thread occupancy by filling each block's available registers well before the available execution units.  Work is ongoing to solve this, but must balance the thread occupancy with increased memory bandwidth requirements from splitting kernels, requiring use of global memory to store intermediate results.

\begin{figure}
    \centering
    \includegraphics[width=\textwidth]{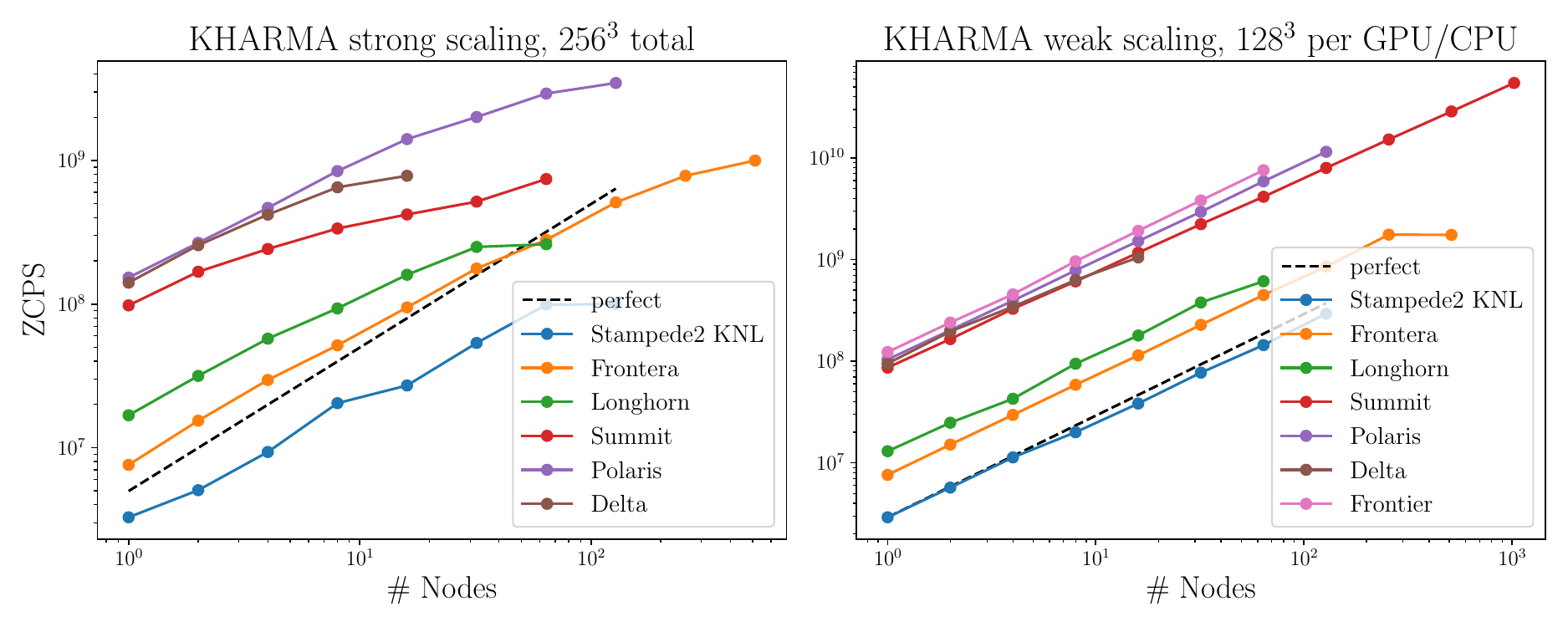}
    \caption{(Left) Strong scaling plot showing \kharma 's performance in ZCPS plotted against the number of GPUs or CPU nodes used, when run on a problem of constant size, $256^3$ zones (scaling results on Delta use a sligtly larger problem, $384^3$ zones). (Right) Weak scaling performance on problems of increasing size, allocating $128^3$ zones per GPU. Scaling results from \cite{prather2022} and personal testing. Results on Delta courtesy of Vedant Dhruv (private communication).}
    \label{fig:kharma_scaling}
\end{figure}

\begin{acknowledgement}
Many thanks to all of the \kharma developers and users, especially Vedant Dhruv, Hyerin Cho, and C\'esar D\'iaz-Blanco, and to Charles Gammie for useful advice, and of course for HARM.
BP gratefully acknowledges support from the Los Alamos National Laboratory Metropolis Fellowship.  Computing time for developing, testing, and applying \kharma has been provided by Oak Ridge Leadership Computing Facility (OLCF) and Argonne Leadership Computing Facility (ALCF) through director's discretionary allocations and through the INCITE program, the National Center for Supercomputing Applications through the Advanced Cyberinfrastructure Coordination Ecosystem: Services \& Support (ACCESS) program, the Texas Advanced Computing Center through the Extreme Science and Engineering Development Environment (XSEDE) and Frontera Large-Scale Community Partnerships (LSCP) programs, and by Los Alamos National Laboratory Institutional Computing.
\end{acknowledgement}

%\section*{Appendix}
%\addcontentsline{toc}{section}{Appendix}
% Don't use \appendix!

\printbibliography

%References should be \textit{cited} in the text by number.\footnote{Please make sure that all references from the list are cited in the text. Those not cited should be moved to a separate \textit{Further Reading} section.} The reference list should be \textit{sorted} in alphabetical order. If there are several works by the same author, the following order should be used: 
%\item all works by the author alone, ordered chronologically by year of publication
%\item all works by the author with a coauthor, ordered alphabetically by coauthor
%\item all works by the author with several coauthors, ordered chronologically by year of publication.
%For the reference style, we suggest to use \textit{LaTeX (US)} from INSPIRE.}

% \begin{thebibliography}
% \cite{LIGOScientific:2016aoc}
% \bibitem{LIGOScientific:2016aoc}
% B.~P.~Abbott \textit{et al.} [LIGO Scientific and Virgo],
% %``Observation of Gravitational Waves from a Binary Black Hole Merger,''
% Phys. Rev. Lett. \textbf{116}, no.6, 061102 (2016)
% doi:10.1103/PhysRevLett.116.061102
% [arXiv:1602.03837 [gr-qc]].
% %7349 citations counted in INSPIRE as of 06 Jan 2022
% \end{thebibliography}

\end{document}